\documentclass[twocolumn]{aastex62}
\usepackage{color}

\newcommand\sersic{S\'{e}rsic}
\newcommand\msun{$M_\odot$}

\defcitealias{KH13}{KH13}

\shortauthors{Zhu et al.}

\begin{document}

\title{The Correlation between Black Hole Mass and Stellar Mass for Classical Bulges and the Cores of Ellipticals}

\author[0000-0002-1333-147X]{Peixin Zhu}
\affiliation{Department of Physics, School of Physics, Peking University, Beijing 100871, China}
\affiliation{Kavli Institute for Astronomy and Astrophysics, Peking University, Beijing 100871, China}

\author[0000-0001-6947-5846]{Luis C. Ho}
\affiliation{Kavli Institute for Astronomy and Astrophysics, Peking University, Beijing 100871, China}
\affiliation{Department of Astronomy, School of Physics, Peking University, Beijing 100871, China}
%\affiliation{Corresponding author}

\author[0000-0003-1015-5367]{Hua Gao}
\affiliation{Department of Astronomy, School of Physics, Peking University, Beijing 100871, China}
\affiliation{Kavli Institute for Astronomy and Astrophysics, Peking University, Beijing 100871, China}
\correspondingauthor{Luis C. Ho}
\email{lho.pku@gmail.com}

\begin{abstract}
The correlation between black hole mass and the stellar mass of the bulge of the host galaxy has attracted much attention ever since its discovery. While traditional investigations of this correlation have treated elliptical galaxies as single, monolithic spheroids, the recent realization that massive elliptical galaxies have undergone significant late-time ($z \lesssim 2$) dissipationless assembly since their initially dense ``red nugget'' phase strongly suggests that black holes in present-day ellipticals should be associated only with their cores and not with their extended envelopes.  We perform two-dimensional image decomposition of Two Micron All Sky Survey $K_s$-band images to derive the stellar mass of the cores of {  35} nearby ellipticals with reliably measured black hole masses. We revisit the relation between black hole mass and bulge stellar mass by combining classical bulges with the cores of ellipticals. The new relation exhibits nearly identical slope {  ($M_{\bullet} \propto M_{\text{core}}^{1.2}$)} as the conventional relation but a factor of $\sim2$ higher normalization and moderately larger intrinsic scatter { (0.4 dex)}.  At a core mass of $10^{11}\,M_{\odot}$, { $M_{\bullet}/M_{\text{core}} = 0.9\%$}, but it rises to {  $M_{\bullet}/M_{\text{core}}=1.5\%$} for the most massive cores with mass $10^{12}\,M_{\odot}$.  { Fast and slow rotator ellipticals follow the same correlation.} The $M_{\bullet}-M_{\text{core}}$ relation provides a revised benchmark for studies of black hole-galaxy coevolution in the high-redshift Universe. 

\end{abstract}


\section{Introduction} \label{sec:1}

Central black holes (BHs) are an integral by-product of the formation and evolution of galaxies, one that may have far-reaching implications for their lifecycle.  All massive galaxies with bulges host BHs with $M_{\bullet} \approx 10^6-10^{10}\,M_\odot$ (\citealt{KH13} and references therein), and it now appears that sub-$10^6\,M_\odot$ BHs occupy even a significant fraction of dwarf and late-type galaxies \citep{gre20}.  Much attention has been devoted to delineating empirical correlations between BH mass and the properties of their host galaxy bulge, especially luminosity \citep{KR95}, stellar mass \citep[e.g.,][]{mag98,har04,sa16}, and stellar velocity dispersion \citep[e.g.,][]{tremaine02,gul09,mcc13}.  Since its original discovery \citep{ferrarese00,gebhardt00}, the $M_{\bullet}-\sigma_\star$ relation has garnered most observational and theoretical interest because it was purported to exhibit the lowest intrinsic scatter and hence was regarded as physically most fundamental.  However, subsequent scrutiny drew attention to the fact that the relation is tight only for a subset of galaxies, namely the ellipticals and disk galaxies possessing classical bulges.   Pseudo bulges, the product of secular evolution \citep{KormendyKennicutt2004}, depart markedly from the $M_{\bullet}-\sigma_\star$ relation of ellipticals and classical bulges in terms of their scatter and zeropoint (\citealt{hu08}; \citealt{gre10}; \citealt{kormendy11}; \citealt{KH13}).  Closer examination by \cite{KH13} of the extant collection of the most reliable dynamically detected BH masses revealed that for classical bulges and ellipticals the $M_{\bullet}-M_{\text{bulge}}$ relation formally has the same intrinsic scatter ($\epsilon = 0.28$ dex) as the $M_{\bullet}-\sigma_\star$ relation.  This result has important implications not only for understanding the physical basis of the BH-galaxy scaling relations, but, from a pragmatic point of view, it sharpens $M_{\text{bulge}}$ as a viable, alternative tool to investigate BH host galaxies, especially under circumstances when stellar kinematics are tough or impractical to access.

From the traditional perspective of galaxy morphology, the entirety of an elliptical galaxy is usually considered a single ``bulge.''  Indeed, previous studies of the $M_{\bullet}-M_{\text{bulge}}$ relation implicitly regarded ellipticals as single, monolithic objects.  However, progress over the past decade has painted a much more complex story for the evolutionary history of elliptical galaxies, a history that stretches from their rapid, early formation epoch to their subsequent gradual assembly and morphological transformation over the past 10 Gyr ($z \lesssim 2$).  At high redshift, the likely progenitors of present-day massive ellipticals---often termed ``red nuggets''---are found to be significantly more compact than their local descendants \citep[e.g.,][]{daddi05,trujillo06,dam11}.  From $z \approx 2$ to $z \approx 0$, red nuggets puffed up in size by a factor of $\sim 3-5$ and roughly doubled in stellar mass \citep[e.g.,][]{buitrago08,van10,van14,davari17}. Several physical explanations have been proposed to account for such dramatic size growth, including dry minor mergers \citep{na07}, mergers plus selection effects \citep{van09}, and adiabatic expansion due to feedback from active galactic nuclei \citep{fan08}. Among these possibilities, the two-phase formation scenario \citep{os10,johansson12} has emerged as most promising in accounting for a number of observational constraints \citep{bez09,hopkins09,hu13, hu13b,hu16, dso14, gre19, oya19} and numerical simulations \citep{kho11,gab12,wel16}. In this picture, the early evolutionary phase of massive galaxies was dominated by dissipative, gas-rich major mergers and cold-phase accretion that led to efficient in situ star formation. This first phase lasted until $z\approx 2$, at which point rapid gas consumption coupled with feedback effects left behind a compact, dense and quenched red nugget.  The second phase of the evolution, which ensued over the subsequent $\sim 10$ Gyr, accumulated an extended envelope through a series of dissipationless, dry minor mergers.  

The two-phase formation history of elliptical galaxies leaves a detectable imprint on the photometric structure of nearby systems.  Analyzing high-quality optical images from the Carnegie-Irvine Galaxy Survey \citep{ho11}, \citet{hu13} found that local massive ($M_{\star} \geqslant1.3 \times 10^{11}\,M_{\odot}$) ellipticals generically contain three photometric sub-structures: a compact inner component with $R_e \lesssim 1$ kpc, an intermediate-scale middle component with $R_e \approx 2.5$ kpc, and an extended outer envelope with $R_e \approx 10$ kpc. \citet{hu13b} argued that the combined structure of the two innermost components (hereinafter the ``core") bears strong similarity to the high-$z$ red nuggets, both in terms of their mass--size relation and median stellar mass surface density profile.  These findings have been extended by subsequent observational work using larger \citep{dso14,del16,oh17} and deeper \citep{spa17} samples, bolstered by cosmological numerical simulations \citep{coo13,coo15,wel16,pil18}.
	
The above picture for the formation and assembly of giant ellipticals leads to certain unavoidable implications for the connection between the mass of the central BH and the stellar mass of the host.  Present-day giant ellipticals universally host the most massive BHs known, with $M_{\bullet} \gtrsim 10^9$ \msun\ (\citealt{KH13}; \citealt{meh19}; \citealt{Liepold20}).  These supermassive BHs gained most of their mass through episodes of vigorous accretion when they shone as distant ($z \approx 2-7$) quasars.  The host galaxies of the high-redshift quasars, although poorly constrained in terms of direct observations of their stellar properties \citep{pen06,mechtley12,marian19}, cannot be any other than the forebearers of today's massive ellipticals and, as a corollary, the immediate predecessors of the red nuggets.  While massive galaxies grew substantially in stellar mass and even more in size since $z \approx 2$, it is widely acknowledged that much of growth occurred dissipationlessly, for massive galaxies have largely remained red and dead after being quenched \citep{ds16}.  And so, too, their central BHs, for the downsizing of active galactic nuclei closely mirrors the cosmic downsizing of star formation \citep{Kriek08,labita09}.  An immediate deduction from these observations is that the central BHs of massive ellipticals must be largely already in place by $z\approx 2$, their masses essentially ``locked in'' to their present-day values in the red nuggets.  The subsequent mass growth of red nuggets occurred through dissipationless accretion of ex situ stars associated with less massive galaxies.  While some of the accreted galaxies themselves may host low-mass BHs \citep{gre20}, these BHs constitute a tiny fraction of the total stellar mass and would have added little to the final weight of the central BH in the massive primary galaxy.  This has two consequences: (1) the ratio of BH mass to total stellar mass decreases toward lower redshifts, and (2) the central BH mass should be related more physically to the ``core'' stellar mass (that which can be traced directly to the red nugget phase) instead of the total stellar mass of present-day ellipticals.  The first issue was discussed in Kormendy \& Ho (2013; their Section 8.6.7).  This paper addresses the second issue.

The preceding considerations motivate us to revise the conventional $M_{\bullet}-M_{\text{bulge}}$ correlation, focusing, in particular, on the population of ellipticals that occupy the upper end of the BH and galaxy mass distribution.  Under the assumption that the inner cores of $z\approx 0$ giant ellipticals provide a faithful fossil record of the stellar component associated with the once-active central BH, we aim to decompose the central cores of all nearby ellipticals hosting central BHs that are detected robustly by dynamical techniques. Utilizing near-infrared images to estimate stellar masses, we redefine a new scaling relation---dubbed the $M_{\bullet}-M_{\text{core}}$ relation for clarity---using the core masses of the ellipticals in conjunction with the sample of classical bulges investigated by \cite{KH13}.  We lump classical bulges together with the cores of ellipticals because we regard them as sharing similar formation physics.

Section~\ref{sec:2} introduces the sample and data. Section~\ref{sec:3} describes the photometric decomposition.  Estimates of the core masses and the new $M_{\bullet}-M_{\text{core}}$ relation are presented in Section~\ref{sec:4}.  Section~\ref{sec:5} discusses the implications of our results, { and a brief summary is given in Section~6}.  For consistency with \cite{KH13}, upon which most of our data draw, we assume the following parameters from the WMAP five-year cosmology \citep{komatsu09}: \mbox{$H_0 = 70.5\,{\rm km~s^{-1}~Mpc^{-1}}$}, \mbox{$\Omega_m = 0.27$}, and \mbox{$\Omega_{\Lambda} = 0.73$}.

\section{Sample and Data} \label{sec:2}

Our analysis focuses on the elliptical galaxies with robust dynamical BH masses from \cite{KH13}.  Of the 46 ellipticals investigated by \cite{KH13}, these authors excluded 17 for different scientific reasons (five for being ``mergers-in-progress,'' nine suspected of having their BH masses compromised by large ionized gas velocity dispersion, and three for hosting an apparently overmassive BH). We follow Kormendy \& Ho's criteria in judging which galaxies have reliable BH masses, except that we reintroduce the two most massive galaxies (NGC 3842 and NGC 4889), which give us the most leverage in anchoring the high-mass end of the sample. While these two galaxies have overmassive BHs with respect to the $M_\bullet-\sigma_\star$ relation, they do not deviate substantially from the $M_{\bullet}-M_{\text{bulge}}$ relation (Figure~16 of \citealt{KH13}). We supplement our sample with seven ellipticals with BH mass measurements published after \cite{KH13}: \mbox{NGC 2974}, \mbox{NGC 4552}, \mbox{NGC 4621}, \mbox{NGC 5813}, and \mbox{NGC 5846} from \citet{sa16}, { \mbox{NGC 1600} from \citet{tho16},} and \mbox{Holm 15A}, the galaxy with the most massive BH known ($M_\bullet =4.0\times10^{10}\,M_\odot$; \citealt{meh19}).   The 38 ellipticals---the main subject of this study---are summarized in the first portion of Table~\ref{tab:1}.

\cite{KH13} stress that BH mass correlates tightly only with the velocity dispersion and stellar mass of ellipticals and the classical bulges of spiral and S0 galaxies.  Pseudo bulges, which evolve through internal secular processes, do not participate in these fundamental scaling relations.  For the purposes of reevaluating the $M_{\bullet}-M_{\text{bulge}}$ relation involving the cores of ellipticals, we extend the dynamic range of the parameter space using the 17 classical bulges considered by \cite{KH13} to have reliable BH masses.  The classical bulges will be compared in their entirety with the cores of ellipticals and thus do not require any decomposition beyond those already provided by \cite{KH13}.  For convenience, we include these 17 classical bulges in the lower portion of Table~\ref{tab:1}.  The final sample, therefore, contains 55 galaxies (38 ellipticals and 17 disk galaxies with classical bulges).

Our primary goal is to derive the stellar mass of the inner core component of the ellipticals.  To this end, we perform multi-component two-dimensional (2D) image decomposition in a manner similar to \citet{hu13}, but instead of optical images we use near-infrared images from the Two Micron All Sky Survey \citep[2MASS;][]{skr06}.  We concentrate on the longest wavelength \mbox{$K_s$} band for the following reasons: (1) it minimizes dust extinction; (2) it is most sensitive to the dominant old stellar population; (3) using the empirical prescription of \cite{KH13}, the mass-to-light ratio in this band, and hence stellar mass, can be estimated reasonably well with knowledge of the stellar velocity dispersion, even in the absence of color information, which is otherwise difficult to secure for the inner cores of these galaxies; (4) 2MASS data are available uniformly for our sample; and (5) although 2MASS images have relatively shallow depth and a mediocre point-spread function (PSF; full width at half maximum FWHM $\approx 3$\arcsec), they suffice for our applications.  An external comparison of the 2MASS photometry is given in \mbox{Appendix~\ref{sec:Appa}}.

We use 2MASS images generated by the image mosaic tool {\tt Montage}\footnote{\tt http://montage.ipac.caltech.edu/} for all galaxies except M87, whose image was taken from the 2MASS Large Galaxy Atlas \citep[][]{jar03}.   We choose a size of $720''\times720''$ for the mosaic images, which comfortably covers the full extent of the galaxies while still leaving sufficient surrounding area for a robust estimation of the sky background.  M87 is more extended, and for its Large Galaxy Atlas image we choose a size of $1200''\times1200''$.  The pixel scale of all images is 1\arcsec\ after drizzling.  The exposure time is 1.3~s. 

\bigskip
\bigskip

\section{Photometric Decomposition} \label{sec:3}

We use {\tt GALFIT} \citep{pe02,pe10} to model the 2MASS $K_s$-band images of the 38 ellipticals to extract photometric parameters of their cores. We generally follow the strategy outlined by \citet{hu13} for their decomposition of optical images of nearby ellipticals, making allowances for certain adjustments necessary for the shallower and lower resolution 2MASS images.

\subsection{Preliminary Steps}

We begin by generating a source mask using {\tt SExtractor} \citep{bertin96}, which identifies all sources in the images and classifies them into stellar and nonstellar objects. Setting the detection threshold to 3~$\sigma$ above the background is generally effective, but this misses some faint stars that are superposed on the main body of the galaxy of interest. We visually examine each image and manually include additional objects to the segmentation image if necessary.  The conservative detection threshold we choose does not always fully mask the faint, outer halos of bright sources. Under these circumstances we enlarge the mask by convolving the segmentation image with an appropriate Gaussian kernel. The final adopted mask is large enough to minimize the contamination from unrelated sources while still leaving enough pixels to create a robust model of the sky background.

The sky background of 2MASS images is complicated by the presence of a large-scale gradient that needs to be removed carefully, a step not yet applied to the mosaic images.  We fit the unmasked sky pixels using a second-order polynomial to remove the global sky level.  To estimate the uncertainty of the sky, which arises principally from residual large-scale fluctuations, we randomly sample the sky pixels and study the variance of the sky distribution.  We repeat the process using boxes of different size to determine the dependence of sky variance on box size.  The sky variance converges on scales $\gtrsim 30$\arcsec, and we use a final box size of $30''\times30''$ to determine the sky uncertainty ($\sigma_{\rm sky}$).  A detailed description of this method, which is based on \citet{li11} and \citet{hu13}, can be found in Appendix B of \citet{gao17}.

To generate the PSF image required by {\tt GALFIT}, for each science image we choose five isolated, bright, unsaturated stars, which are combined using standard {\tt IRAF} procedures to produce a PSF of high signal-to-noise ratio.  The final PSFs have FWHM $\approx$ 2\farcs8$-$3\farcs2, consistent with the typical seeing of 2MASS.

\bigskip
\bigskip

\subsection{Model Fitting}

With the mask, PSF, and sky-subtracted science images in hand, we proceed to model fitting using {\tt GALFIT}.  As in \citet{hu13}, we start with the simplest single-component fit and, as the data allow, gradually increase the complexity of the model by adding extra components.  The radial surface brightness of each component is represented with the \citet{sersic68} function, which is flexible enough to accommodate most galaxy structures and can be compared easily with published results:

\begin{equation}
\mu(R)=\mu_e\exp \left\{-\kappa\left(\left(\frac{R}{R_e}\right)^{1/n}-1\right)\right\}.
\end{equation}

\noindent
The free parameters are the \sersic{} index $n$ describing the profile shape, the effective radius $R_e$, surface brightness $\mu_e$ at $R_e$, centroid $x_0$ and $y_0$, ellipticity $e$, and position angle PA. The parameter $\kappa$ depends on $n$. Instead of including an additional component to account for the sky, as in \citet{hu13}, we simply fix the sky value to zero because our images are already sky-subtracted.  

As with \citet{hu13}, we find that our models, even those with multiple components, are not very sensitive to the choice of initial parameters. Thus, we generally set the same initial parameters for all the galaxies. Special treatment is necessary only when certain models yield obviously unphysical parameters (e.g., unrealistically large $R_e$, or very small values of $n$ and $e$ that are clearly inappropriate for elliptical galaxies), or when {\tt GALFIT} gets stuck in a global minimum solution and crashes.  Under these circumstances, some adjustment of the initial parameters is needed for the fit to converge.  The quality of the model often cannot be judged solely by formal statistical criteria.  Each fit needs to be evaluated based on the residuals of the one-dimensional radial profile of surface brightness, model-data consistency of structural parameters such as $e$ and PA, and detailed inspection of the residual image.

\subsection{Best-fit Parameters and their Uncertainties}\label{sec:3.3}

We obtained reliable fits for 37 out of the 38 ellipticals: one requires a four-component model, 23 are well-described by three-component models, 11 only needs models with two components,  and two suffice with just a single component.  The only exception for which we could not obtain a useful fit is \mbox{NGC 2974}, which is severely contaminated by a nearby bright star. We exclude this galaxy from further consideration.  

\begin{figure*}[htb!]
\epsscale{1}
\plotone{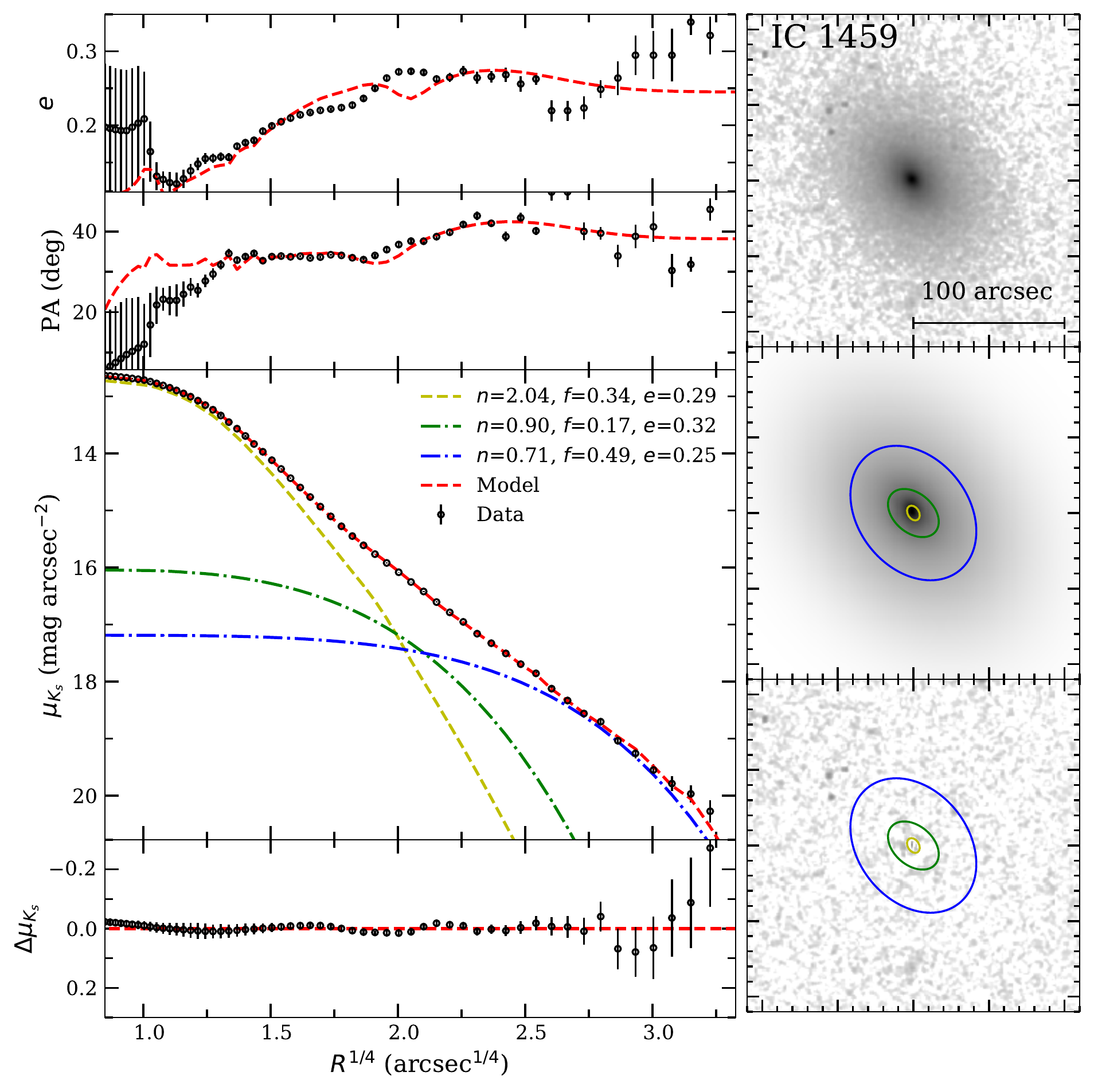}
\caption{The best-fit three-component model of IC 1459. The right panels display, from top to bottom, images of the data, the best-fit model, and the residuals. All images are shown using the same logarithmic stretch, and they are centered on the galaxy centroid, with north up and east to the left.  The left panels display the isophotal analysis of the 2D image fitting. From top to bottom, the panels show radial profiles of the ellipticity ($e$), the position angle (PA), the $K_s$-band surface brightness ($\mu_{K_s}$), and the fitting residuals ($\bigtriangleup\mu_{K_s}$).  Profiles of the data, the model, and the individual components are encoded consistently with different symbols, line styles, and colors, as explained in the legend.  \\(The complete figure set of 37 images is available online.)
\label{fig:1}}
\end{figure*}

As an example, \mbox{Figure~\ref{fig:1}} illustrates the best-fit three-component model for \mbox{IC 1459}. The full set of models for the entire sample is provided in the electronic version of the paper.  The parameters for all the best-fit models appear in Table~\ref{tab:2}.  \mbox{Appendix~\ref{sec:Appb}} gives a detailed explanation of the galaxies that were fit with non-standard models (with fewer or greater than the usual three components).  The inner, middle, and outer components of IC~1459 account for $f = 34\%$, 17\%, and 49\% of the flux. If we consider the inner plus middle components as the core, then the core comprises 51\% of the total flux.  For the sample as a whole, the median fractional contribution of the core to the total flux is 64\%.

The uncertainties of the best-fit parameters arise from two main contributions, the formal statistical uncertainty from {\tt GALFIT} and the systematic uncertainty stemming from the influence of the PSF and sky subtraction.  For ground-based observations of nearby, inactive galaxies, systematic effects due to sky subtraction usually dominate the error budget \citep{guo09,yoo11,hu13}.  For the vast majority of the objects in our sample, the innermost component is quite well-resolved and therefore not critically dependent on precise knowledge of the PSF.  The inner component of our galaxies has a median $R_e \approx 0.8$ kpc, which corresponds to an angular scale of 7\farcs1 at a typical sample distance of 25 Mpc, or $\sim 5$ times the half width at half maximum of the 2MASS PSF (1\farcs5).  To evaluate quantitatively the impact of possible variations in the PSF, we repeat the fits with alternate versions of the PSF rendered from different combinations of bright stars.  To evaluate the effect of sky subtraction, we redo the fits after perturbing the sky value by $\pm 1\, \sigma_{\rm sky}$.  The final uncertainty is the quadrature sum of the statistical and systematic uncertainties.  On average, the relative uncertainties are quite small (Table~\ref{tab:2}). The parameters most vulnerable to sky error are $R_e$ and $n$.  For the inner, middle, and outer components, the mean uncertainties for $R_e$ are, respectively, $\sim6\%$, $3\%$, and $4\%$; for $n$, the corresponding values are $\sim4\%$, $9\%$, and $8\%$. The mean uncertainties for all parameters in this study are smaller than those in \citet{hu13}, principally because our larger image size enables the sky background to be better determined.

\section{Results}\label{sec:4}

\subsection{Stellar Mass of the Core} \label{sec:core}

To construct the $M_{\bullet}-M_{\text{core}}$ correlation, we need to obtain the stellar mass of the cores of the ellipticals.  As discussed in Section~\ref{sec:1}, \citet{hu13,hu13b} provide compelling evidence that the inner plus middle components of the three-component decomposition constitute the remnants of high-redshift red nuggets.  We designate the sum of the inner and middle components as the ``core'' of present-day ellipticals.  For the minority of galaxies whose best fit only requires two components, Appendix~\ref{sec:Appb} shows that the smaller of the two components is equivalent to the unresolved sum of the inner and middle components of the standard three-component substructure.  Thus, we similarly regard as ``core'' the inner component of the two-component galaxies. { Our new relation excludes NGC~4486A and Holm~15A, the two galaxies for which only a single \sersic\ component is required (see  Appendix~\ref{sec:Appb}).}

As discussed in Kormendy \& Ho (2013; their Section 6.6.1), the $K$-band stellar mass-to-light ratio can be predicted from an empirical correlation between $M/L_K$ and either the optical color or stellar velocity dispersion.  We do not have adequate information on optical colors, but fortunately stellar velocity dispersions are available (Table~\ref{tab:1}).  From Equation 8 of \cite{KH13}, the stellar mass for the core follows from

\begin{equation}
\log M_{\rm core}/L_{K_s}=0.2871 \log \sigma_\star-0.6375;\ \ \text{rms}=0.088,
\end{equation}

\noindent
with $M_{\rm core}$ the stellar mass and $L_{K_s}$ the luminosity of the core in solar units and $\sigma_\star$ the stellar velocity dispersion in ${\rm km\,s^{-1}}$  measured at $R_e$.

\subsection{The $M_{\bullet}-M_{\text{bulge}}$ and $M_{\bullet}-M_{\text{core}}$ Relations} \label{sec:relation}

In light of the significant increase in the number of elliptical galaxies with BH mass measurements, we first offer an update of the conventional $M_{\bullet}-M_{\text{bulge}}$ relation.  Here $M_{\text{bulge}}$ pertains to classical bulges and the integrated light of the entire elliptical galaxy.  As in \cite{KH13}, we exclude pseudo bulges because of their different formation channel.  We perform a linear regression using {\tt LinMix} (\citealt{kel07}; Python package by \citealt{mey15}), which allows for the inclusion of measurement errors in both variables while still accounting for a component of intrinsic, random scatter.  The revised  relation (Figure~\ref{fig:2}a),

{
\begin{equation}
\frac{M_{\bullet}}{10^9\,M_{\odot}}=(0.47^{+0.06}_{-0.05})(\frac{M_{\text{bulge}}}{10^{11}\,M_{\odot}})^{1.18\pm0.08},
\end{equation}
}

\noindent
is very similar to the fit of \cite{KH13} in terms of normalization and slope. { Our intrinsic scatter, $\epsilon = 0.34$ dex, is also nearly identical to that of Kormendy \& Ho (2013; $\epsilon = 0.28$ dex).}

Combining classical bulges with only the {\it cores}\ of ellipticals, we arrive at a new $M_{\bullet}-M_{\text{core}}$ relation (Figure~\ref{fig:2}b):

{
\begin{equation}	
\frac{M_{\bullet}}{10^9\,M_{\odot}}=(0.90^{+0.14}_{-0.12})(\frac{M_{\text{core}}}{10^{11}\,M_{\odot}})^{1.22\pm0.11}.
\end{equation}
} 

\noindent
Compared to the $M_{\bullet}-M_{\text{bulge}}$ relation, the slope of the $M_{\bullet}-M_{\text{core}}$ relation is nearly identical, but the normalization is a factor of $\sim$2 higher, and the intrinsic scatter is moderately larger { ($\epsilon = 0.41$ dex)}.  Since the relation is slightly super-linear, the relative BH mass fraction depends mildly on the stellar mass of the core. Rewriting Equation (4) in a different form,

{
\begin{equation}
100\frac{M_{\bullet}}{M_{\text{core}}}=(0.90^{+0.14}_{-0.12})(\frac{M_{\text{core}}}{10^{11}\,M_{\odot}})^{0.22\pm0.11}.
\end{equation}
}

\noindent 
At a core mass of $10^{11}\,M_{\odot}$, the BH mass fraction is { $M_{\bullet}/M_{\text{core}} = 0.9\%$,} rising to { $M_{\bullet}/M_{\text{core}}=1.5\%$ }for the most massive cores with $M_{\text{core}} = 10^{12}\,M_{\odot}$.

\section{Discussion}\label{sec:5}

{\subsection{Cores of Local Ellipticals as Relics of Red Nuggets}}

The empirical correlations between BH mass and bulge properties have provided a useful framework for investigating the connection between BH growth and galaxy evolution.  Much effort has been invested in characterizing the BH-host galaxy scaling relations, both for active and inactive galaxies, from the local Universe (\citealt{KH13}) to the highest redshift quasars \citep{decarli18,izumi19}.  Any discussion of the possible cosmic evolution of the scaling relations must be anchored on a proper $z = 0$ reference.  Which version of the myriad available BH-host relations should serve as the local standard?  The answer depends on the scientific context in hand.

In the case of the $M_{\bullet}-M_{\text{bulge}}$ relation, in addition to excluding pseudo bulges and other objects based on physical or experimental grounds \citep{KH13}, we emphasize that thought needs to be given to consider which portion of an elliptical galaxy should be included in $M_{\text{bulge}}$.  In light of the abundant evidence for the two-phase formation history of massive ellipticals, we argue that the traditional, tacit assumption that the entire galaxy belongs to the ``bulge'' is inappropriate.  Instead, we suggest that we include only that portion of the structure that participated in the coeval evolution of the central BH during the quasar era.  The relevant component in local ellipticals is their ``core'', which is the sum of the inner and middle components in the framework of the three-component decomposition of \citet{hu13}. The cores of ellipticals are none other than the remnants of the so-called red nuggets at $z \approx 2$, whose progenitors hosted quasars that were responsible for much of the BH growth.  Among the arguments advanced by \citet{hu13b} to link the inner structure of local ellipticals to the red nugget population is the fact that the combination of the inner plus middle components follows a tight stellar mass-size relation very similar to that observed in high-$z$ early-type galaxies (ETGs).  The same argument holds for our sample.  Figure~\ref{fig:3} compares the cores of our ellipticals to the ETGs at $1 \lesssim z \lesssim 3$ studied by \citet{van14}.  As the sizes from \citet{van14} are given in the $V$ band, we convert their mass-size relation to the $K_s$ band using the average size gradient of $\Delta \log R_e/\Delta \log \lambda = -0.25$, which for ETGs does not seem to depend much on mass and redshift.  This comparison shows that the cores of our sample of nearby ellipticals closely follow the mass-size relation of high-$z$ ETGs.  { Figure~3 additionally plots the individual measurements for the inner, middle, and outer components of the galaxies in our sample that were decomposed successfully with a three-component model.  As emphasized in Huang et al. (2013a, their Figure~28) and Huang et al. (2013b, their Figure~1b), it is striking that each of the subcomponents obeys a separare, relatively well-defined mass-size relation, which lends credence to the notion that the photometric decomposition indeed extracts physically meaningful substructures.  Our present analysis, although based on images of lower resolution and signal-to-noise ratio than those enjoyed by \citet{hu13}, nevertheless reproduces the essential trends presented by those authors.}

{\subsection{Fast versus Slow Rotators}}

\begin{figure*}[tb!]
\epsscale{1.2}
\plotone{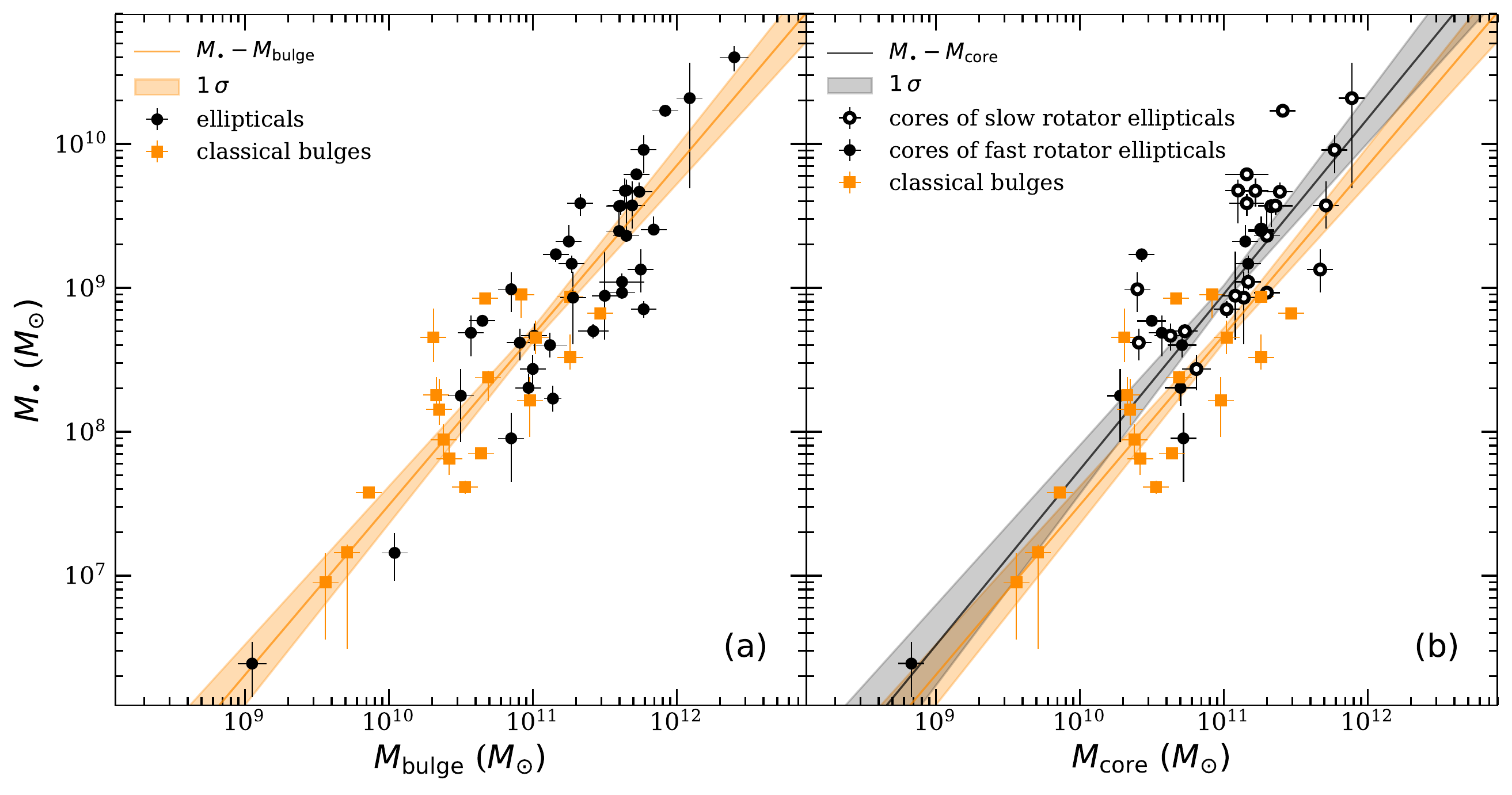}
\caption{Correlation between BH mass ($M_{\bullet}$) and (a) total stellar mass ($M_{\text{bulge}}$) of classical bulges and ellipticals and (b) stellar mass of classical bulges and the core mass ($M_{\text{core}}$) of ellipticals. {  Elliptical galaxies are divided into fast rotators (solid circles) and slow rotators (open circles).} Best-fit relations (solid lines) and their 1~$\sigma$ scatter (shaded areas) are shown.  
\label{fig:2}}
\end{figure*}

\begin{figure*}[tb!]
\epsscale{0.8}
\plotone{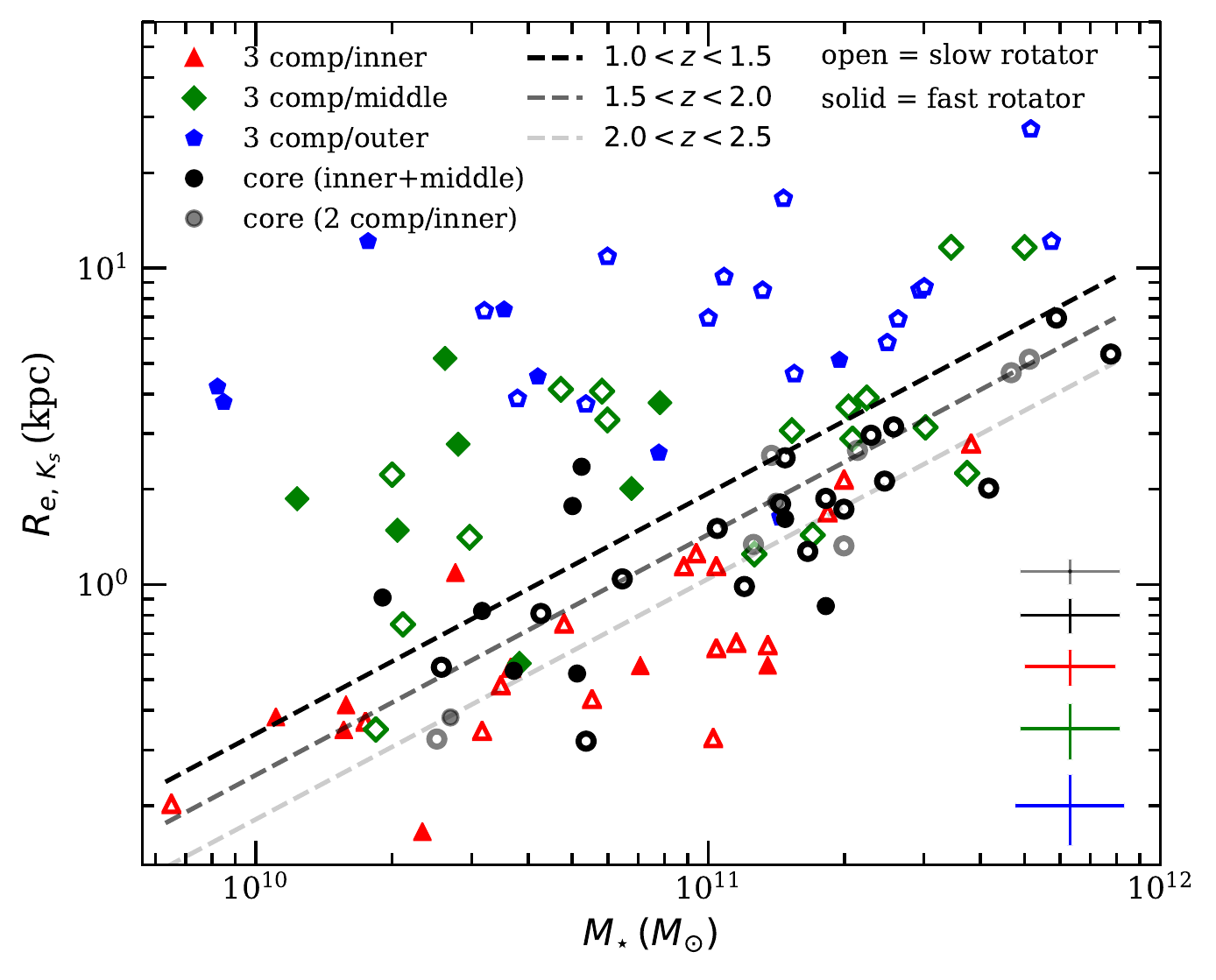}
\caption{{ Mass-size relation for high-$z$ ETGs (dashed lines give the best-fit relations for different redshift ranges; \citealt{van14}) and for the cores of nearby elliptical galaxies derived in this study. We show the measurements separately for the inner (red), middle (green), and outer (blue) components, as well as for combination of the inner and middle components, which we designate as the ``core'' (black). For the few objects that were fit with only two components, the cores (grey) are simply the  more inner of the two components (see Appendix~\ref{sec:Appb}).  Slow rotators and fast rotators are denoted by the open and solid symbols, respectively. Typical uncertainties are given in the lower-right corner.}
\label{fig:3}}
\end{figure*}

{ A fundamental thesis of our paper is that the remnants of the stellar progenitors that once hosted the most massive, actively growing BHs at high redshifts can be discerned as photometrically distinct central substructures in local elliptical galaxies.  We identify these substructures as the inner and middle components of our multi-component decomposition, which we combine and designate as the ``core.''  How robust is this assumption, in view of the diverse formation channels of elliptical galaxies?  As summarized in \cite{kor09}, ETGs fall into two broad classes: (1) luminous, giant ellipticals that have relatively flat, core nuclear profiles, boxy isophotes, and slow rotation; (2) less luminous members that display cuspy, coreless central profiles, disky isophotes, and fast rotation.  The boundary between the two\footnote{To avoid confusion with our usage of the term ``core'' to represent to the central compact structure of ellipticals (inner plus middle components from our decomposition), hereinafer we refer to the two ETG classes as ``slow rotators'' and ``fast rotators,'' which are closely connected to the core and coreless nuclear profile classifications, respectively \citep{emsellem11}.} roughly lies at $M_V \approx -21.5$~mag, which corresponds to $M_* \approx 2\times10^{11}\,M_\odot$.  The distinctive photometric and kinematic properties of the two classes of ETGs reflect differences in their evolutionary pathway \citep{cap16}.  Dissipative processes such as wet mergers and gas accretion likely play a key role in the formation of fast rotators, whereas dissipationless, dry mergers more strongly shape slow rotators.  However, debate lingers as to whether the two populations share the same progenitors at $z \gtrsim 2$.  As reviewed by \cite{cap16}, recent observations suggest that present-day fast rotators were once gas-accreting, star-forming disks that became quenched by growing bulges, to be contrasted with slow rotators, which gained mass rapidly at early times and then later puffed up mainly through dry merging.  This picture appears to be at odds with current cosmological numerical simulations, which find that the progenitors of fast and slow rotators are indistinguishable at $z>1$ \citep{pen17}.  This ambiguity calls into question the assumption that the cores of present-day fast and slow rotators map onto the same set of compact progenitors at $z \approx 2$.  A study of a small sample of massive galaxies selected from the Illustris simulation shows that only $\sim50\%$ of compact ETGs at $z\approx2$ survive as intact cores of $z\approx 0$ ellipticals \citep{wel16}.  There is also no concensus on the relative importance of major and minor mergers.  For instance, \cite{KH13} contend that major mergers contribute to the formation of both fast and slow rotators, while discourse on the mass-size relation places greater emphasis on minor mergers (e.g., \citealt{bez09}; but see \citealt{davari17}).  

Notwithstanding these complications, there is little doubt that the overall framework of the two-phase formation scenario has significant merit for interpreting ETGs.  Both observations and simulations show that, despite the differences between fast and slow rotators inside $R_e$ or the complexities associated with their interpretation, the outer envelopes across the whole ETG population share a common origin: late-time accretion of ex situ stars, principally through minor mergers \citep{coo13,coo15,na14,hu16,rodriguez16,wel16,pen17,pul18,pul20}.  As such, we reassert the central tenet of this study---the outskirts of local ellipticals should {\it not}\ be included in the final tally of the ``bulge'' stellar mass when considering its scaling relation with the BH mass.

As nuclear profiles are available for the entire sample (Table~1), we can investigate the possible dependence of our results on ETG class.  Figure~\ref{fig:2} separately highlights the distribution of fast and slow rotators on the $M_{\bullet}-M_{\text{bulge}}$ and $M_{\bullet}-M_{\text{core}}$ relations.  Apart from the expected dependence on stellar mass, no other trend emerges.  Interestingly, the cores of slow rotators follow a tight mass-size relation (slope $\alpha=0.89\pm0.10$; scatter $\sigma=0.14$ dex) that closely resembles that of $z\gtrsim 2$ ETGs \citep{van14}.  Fast rotators appear to exhibit larger scatter ($\sigma \approx 0.4$) and possibly shallower slope ($\alpha \approx 0.4$), but the poor statistics prevent us from drawing any confident conclusions.
}

\bigskip
\bigskip

{
\section{Summary}

In the context of the widely discussed two-phase formation scenario for early-type galaxies, only the cores of elliptical galaxies participated in the initial coeval growth of the central BH.  The outer, extended envelopes of ellipticals, accrued through late-time accretion, should be excluded from the traditional $M_{\bullet}-M_{\text{bulge}}$ relation.  We perform two-dimensional, multi-component decomposition of $K_s$-band images of 35 local ellipticals with reliably measured BH masses to isolate their core structures from their extended outskirts. After estimating the stellar masses of the cores, we use them and the central BH masses to construct a new $M_{\bullet}-M_{\text{core}}$ relation for local elliptical galaxies and classical bulges.}

{ The new relation has nearly identical slope ($M_{\bullet} \propto M_{\text{core}}^{1.2}$) but higher normalization (by a factor of $\sim 2$) and larger intrinsic scatter (by $\sim 24\%$)} compared to the traditional $M_{\bullet}-M_{\text{bulge}}$ relation.  { Fast and slow rotators obey the same relation.} The higher normalization is to be expected, given that the core component comprises $\sim 60\%$ of the total $K_s$-band light (stellar mass).  This has important implications for the interpretation of the BH-host scaling relations for higher redshift active galaxies and quasars, which generally exhibit elevated BH mass fractions \citep[e.g.,][]{pen06,ho07,jah09,wan10,KH13,decarli18,izumi19,pensabene20}.

\begin{figure*}
\epsscale{0.8}
\figurenum{A1}
\plotone{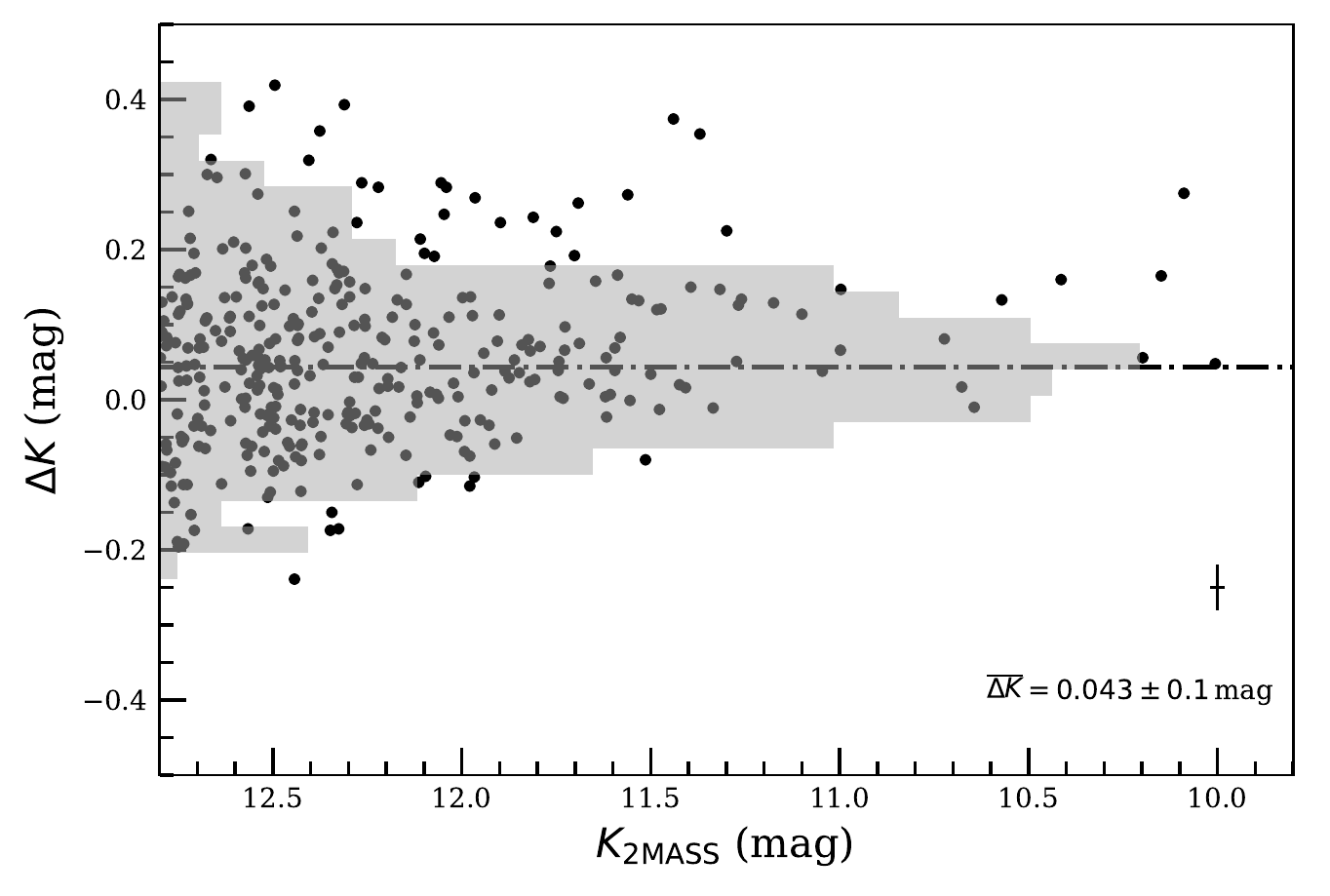}
\caption{Difference in integrated magnitude ($\Delta{K} \equiv K_{\text{WINGS}}-K_{\text{2MASS}}$) of 347 elliptical galaxies as measured in WINGS \citep{mor14} and the 2MASS Extended Source Catalog \citep{jar00}.  Typical uncertainties are shown in the bottom-right corner.  There are no discernible trends with luminosity for galaxies with $K_s<12.8$ mag. The histogram gives the distribution of $\Delta{K}$, which has a mean and standard deviation of $0.043 \pm 0.1$ mag.  \label{fig:A1}}
\end{figure*}

\begin{figure*}
\epsscale{0.8}
\figurenum{B1}
\plotone{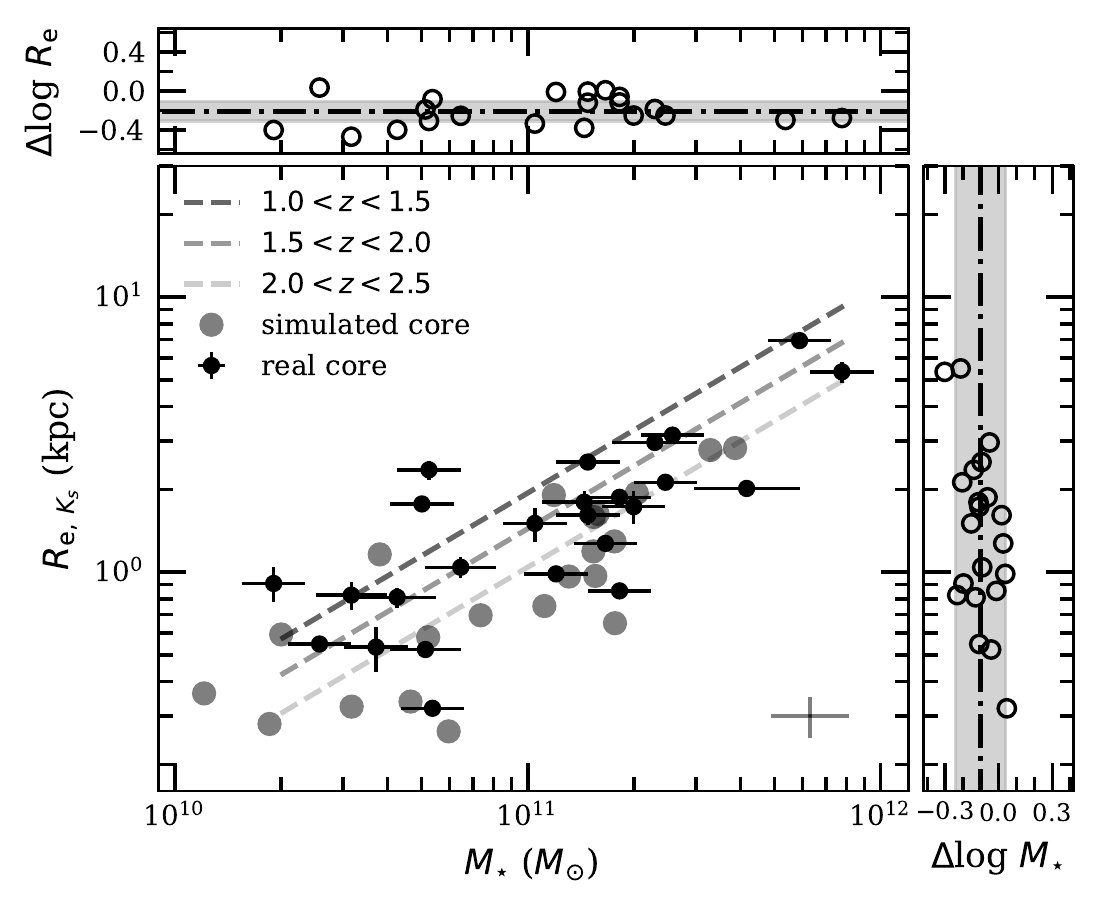}
\caption{Mass-size relation of the cores of nearby ellipticals (black points with error bars) and the cores of simulated galaxies generated by artificially shifting them to larger distances (gray points with typical error bars shown in the bottom-right corner). The dashed lines give the best-fit relations for ETGs observed in different redshift ranges (\citealt{van14}; see Figure 3).  The difference in effective radius is $\log \Delta R_e({\rm simulated-real}) = -0.2 \pm 0.1$ dex (top panel), and the difference in stellar mass is $\log \Delta M_*({\rm simulated-real}) = -0.1 \pm 0.14$ dex (right panel).
\label{fig:B1}}
\end{figure*}

\acknowledgments{This work was supported by the National Science Foundation of China (11721303, 11991052) and the National Key R\&D Program of China (2016YFA0400702).  We made use of {\tt Montage}, funded by the National Aeronautics and Space Administration's Earth Science Technology Office, Computational Technnologies Project, under Cooperative Agreement Number NCC5-626 between NASA and the California Institute of Technology. The code is maintained by the NASA/IPAC Infrared Science Archive. We thank Minghao Guo, Song Huang, Ruancun Li, Jinyi Shangguan, and Jing Wang for advice.  { We are grateful to an anonymous referee for helpful suggestions.}

\appendix

\section{Validation of 2MASS photometry}\label{sec:Appa}

To examine whether the limited depth of the 2MASS images causes flux loss in the extended outskirts of elliptical galaxies, a straightforward method is to compare the 2MASS photometry with that from a deeper survey.  For this purpose, we choose the near-infrared Wide-field Nearby Galaxy-cluster Survey \citep[WINGS;][]{mor14} of 27 nearby galaxy clusters based on data collected with the WFCAM instrument on the United Kingdom Infra-Red Telescope \citep[UKIRT;][]{cas07}.  The survey is 90\% complete in detection rate at total magnitude $K=19.4$\,mag and in classification rate at $K=18.5$\,mag \citep{val09}.  For comparison, 2MASS is complete to $K_s=14.3$\,mag for point sources \citep{skr06}.  In total, 347 elliptical galaxies with \mbox{$K_s<12.8$\,mag} overlap between the two catalogs.  We adopt integrated \citet{kro80} magnitudes from the 2MASS Extended Source Catalog \citep{jar00}. For WINGS, we use their $K$-band MAG\_BEST \citep{val09}.  The UKIRT images use the broad $K$ band of the Mauna Kea photometric system \citep{tok02}, which differs from the $K_s$ band of 2MASS by $K_s - K = 0.004$ mag \citep{car01,bes05}.  Comparison of the two datasets yields $\Delta{K}=K_{\text{WINGS}}-K_{\text{2MASS}} = 0.043 \pm 0.1$ mag (Figure~\ref{fig:A1}), reasonably consistent given the typical photometric uncertainty of 0.03 mag.  This comparison confirms that, while relatively shallow, 2MASS can capture the faint outskirts of elliptical galaxies brighter than $K_s = 12.8$ mag and is adequate for our application.

\section{Notes on objects fit with non-standard models}\label{sec:Appb}

As mentioned in Section~\ref{sec:3.3}, 14 ellipticals in our sample were fit with models other than our standard three-component models: two (\mbox{NGC 4486A} and \mbox{Holm 15A}) were fit with a single component; 11 (\mbox{NGC 3842}, 4291, 4751, 5077, 5328, 5516, 5845, 6086, 6861, 7619, and 7768) were fit with two components; and one (NGC~4697) required as many as four components.  Here we discuss their physical interpretations. 

NGC~4486A is an ultra-compact elliptical galaxy with $R_e=0.56$ kpc, which is less than the typical size of the inner component ($R_e \approx0.8$ kpc). Its compactness precludes us from performing a reliable photometric decomposition with the 2MASS image. Our single-component fit leaves significant residuals, and thus we exclude it from our final analysis. At $D_L\approx250$ Mpc, \mbox{Holm 15A} is the most distant galaxy in our sample. { The seeing of 2MASS is not sufficient to perform detailed photometric analysis of this galaxy, and we exclude it from further consideration. It is worth  noting, however, that the high-resolution imaging studies by \citet{bon15} and \citet{meh19} do show that  \mbox{Holm 15A} is described well by a three-component model.}

The intrinsic structure of NGC 3842 can be described with just two components, with no evidence for a third, extended, outer envelope \citep{but92, sun05}. The same holds for \mbox{NGC 5845}, which \citet{jia12} consider to be a local analog of a $z\approx 2$ red nugget. The two-component flux of these galaxies can be regarded as the integrated light of the core itself.

For the other galaxies treated with two-component models, a more common situation is that the angular size of the galaxy's inner component is too small to be resolved by 2MASS.  This is the case for nearby galaxies with ultra-compact ($R_e\lesssim0.3$ kpc) inner components (\mbox{NGC 4291}, \mbox{4751}, \mbox{5077}, and \mbox{6861}) and more distant ($D_L \gtrsim50$ Mpc) galaxies with moderately compact ($R_e\approx0.8-1$ kpc) inner components (\mbox{NGC 5328}, \mbox{5516}, \mbox{6086}, \mbox{7619}, and \mbox{7768}).  Under such a circumstance, the inner and intermediate components of three-component galaxies blur together into a single core component, and the larger of the two components in the two-component model is simply the outer component (extended envelope) of the normally three-component fit.  We investigated the impact of spatial resolution using, as control, the more nearby galaxies in our sample that are well-resolved into the usual three components.  Following \citet{yu18}, we simulate mock observations of more distant galaxies by artificially moving the images of the nearby galaxies to larger and larger distances.  In general, the original inner component becomes hard to separate from the intermediate component beyond $D_L \approx 70-100$ Mpc, and the best fit reduces to a two-component model consisting of an inner ``core'' plus an outer envelope. Figure~\ref{fig:B1} shows that the core mass and effective radius derived from the simulated images agree reasonably well with the actual values from the real, input images; on average, } $\log \Delta M_*({\rm simulated-real}) = -0.10 \pm 0.14$ dex and $\log \Delta R_e({\rm simulated-real}) = -0.2 \pm 0.1$ dex. Based on these simulations, we conclude that the inner component of the two-component galaxies provides sufficiently accurate estimates of the core mass for the purposes of our main analysis.

\mbox{NGC 4697} is the only galaxy in our sample that requires four components for its fit.  \citet{hu13} reached the same conclusion, suspecting that the galaxy may be a misclassified S0 whose extended envelope consists of two outer components. For the purposes of the current analysis, we simply combine the innermost two components as the core.

\clearpage

\figsetstart
\figsetnum{37}
\figsettitle{Best-fit models of 37 ellipticals}

\figsetgrpstart
\figsetgrpnum{37.1}
\figsetgrptitle{IC 1459}
\figsetplot{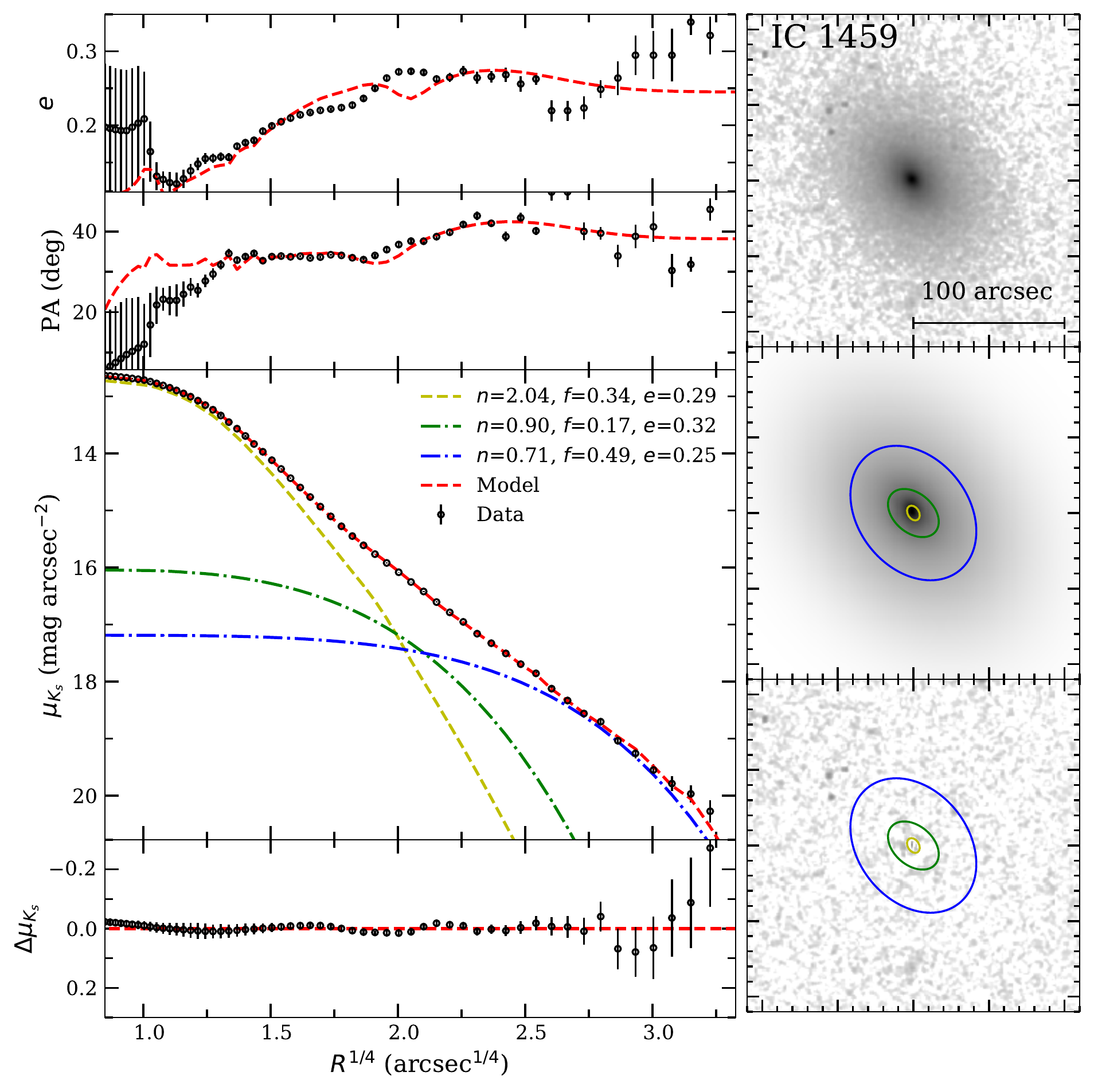}
\figsetgrpnote{The best-fit three-component model of IC 1459. The right panels display, from top to bottom, images of the data, the best-fit model, and the residuals.  All images are shown using the same logarithmic stretch and are centered on the galaxy centroid, with north up and east to the left.  The left panels display the isophotal analysis of the 2D image fitting. From top to bottom, the panels show radial profiles of the the ellipticity ($e$), the position angle (PA), the $K_s$-band surface brightness ($\mu_{K_s}$), and the fitting residuals ($\bigtriangleup\mu_{K_s}$).  Profiles of the data, the model, and the individual components are encoded consistently with different symbols, line styles, and colors, which are explained in the legends.}
\figsetgrpend

\figsetgrpstart
\figsetgrpnum{37.2}
\figsetgrptitle{M32}
\figsetplot{M32.pdf}
\figsetgrpnote{The best-fit three-component model of M32. The right panels display, from top to bottom, images of the data, the best-fit model, and the residuals.  All images are shown using the same logarithmic stretch and are centered on the galaxy centroid, with north up and east to the left.  The left panels display the isophotal analysis of the 2D image fitting. From top to bottom, the panels show radial profiles of the the ellipticity ($e$), the position angle (PA), the $K_s$-band surface brightness ($\mu_{K_s}$), and the fitting residuals ($\bigtriangleup\mu_{K_s}$).  Profiles of the data, the model, and the individual components are encoded consistently with different symbols, line styles, and colors, which are explained in the legends.}
\figsetgrpend

\figsetgrpstart
\figsetgrpnum{37.3}
\figsetgrptitle{NGC 1332}
\figsetplot{NGC1332.pdf}
\figsetgrpnote{The best-fit three-component model of NGC 1332. The right panels display, from top to bottom, images of the data, the best-fit model, and the residuals.  All images are shown using the same logarithmic stretch and are centered on the galaxy centroid, with north up and east to the left.  The left panels display the isophotal analysis of the 2D image fitting. From top to bottom, the panels show radial profiles of the the ellipticity ($e$), the position angle (PA), the $K_s$-band surface brightness ($\mu_{K_s}$), and the fitting residuals ($\bigtriangleup\mu_{K_s}$).  Profiles of the data, the model, and the individual components are encoded consistently with different symbols, line styles, and colors, which are explained in the legends.}
\figsetgrpend

\figsetgrpstart
\figsetgrpnum{37.4}
\figsetgrptitle{NGC 1374}
\figsetplot{NGC1374.pdf}
\figsetgrpnote{The best-fit three-component model of NGC 1374. The right panels display, from top to bottom, images of the data, the best-fit model, and the residuals.  All images are shown using the same logarithmic stretch and are centered on the galaxy centroid, with north up and east to the left.  The left panels display the isophotal analysis of the 2D image fitting. From top to bottom, the panels show radial profiles of the the ellipticity ($e$), the position angle (PA), the $K_s$-band surface brightness ($\mu_{K_s}$), and the fitting residuals ($\bigtriangleup\mu_{K_s}$).  Profiles of the data, the model, and the individual components are encoded consistently with different symbols, line styles, and colors, which are explained in the legends.}
\figsetgrpend

\figsetgrpstart
\figsetgrpnum{37.5}
\figsetgrptitle{NGC 1399}
\figsetplot{NGC1399.pdf}
\figsetgrpnote{The best-fit three-component model of NGC 1399. The right panels display, from top to bottom, images of the data, the best-fit model, and the residuals.  All images are shown using the same logarithmic stretch and are centered on the galaxy centroid, with north up and east to the left.  The left panels display the isophotal analysis of the 2D image fitting. From top to bottom, the panels show radial profiles of the the ellipticity ($e$), the position angle (PA), the $K_s$-band surface brightness ($\mu_{K_s}$), and the fitting residuals ($\bigtriangleup\mu_{K_s}$).  Profiles of the data, the model, and the individual components are encoded consistently with different symbols, line styles, and colors, which are explained in the legends.}
\figsetgrpend

\figsetgrpstart
\figsetgrpnum{37.6}
\figsetgrptitle{NGC 1407}
\figsetplot{NGC1407.pdf}
\figsetgrpnote{The best-fit three-component model of NGC 1407. The right panels display, from top to bottom, images of the data, the best-fit model, and the residuals.  All images are shown using the same logarithmic stretch and are centered on the galaxy centroid, with north up and east to the left.  The left panels display the isophotal analysis of the 2D image fitting. From top to bottom, the panels show radial profiles of the the ellipticity ($e$), the position angle (PA), the $K_s$-band surface brightness ($\mu_{K_s}$), and the fitting residuals ($\bigtriangleup\mu_{K_s}$).  Profiles of the data, the model, and the individual components are encoded consistently with different symbols, line styles, and colors, which are explained in the legends.}
\figsetgrpend

\figsetgrpstart
\figsetgrpnum{37.7}
\figsetgrptitle{NGC 1550}
\figsetplot{NGC1550.pdf}
\figsetgrpnote{The best-fit three-component model of NGC 1550. The right panels display, from top to bottom, images of the data, the best-fit model, and the residuals.  All images are shown using the same logarithmic stretch and are centered on the galaxy centroid, with north up and east to the left.  The left panels display the isophotal analysis of the 2D image fitting. From top to bottom, the panels show radial profiles of the the ellipticity ($e$), the position angle (PA), the $K_s$-band surface brightness ($\mu_{K_s}$), and the fitting residuals ($\bigtriangleup\mu_{K_s}$).  Profiles of the data, the model, and the individual components are encoded consistently with different symbols, line styles, and colors, which are explained in the legends.}
\figsetgrpend

\figsetgrpstart
\figsetgrpnum{37.8}
\figsetgrptitle{NGC 1600}
\figsetplot{NGC1600.pdf}
\figsetgrpnote{The best-fit three-component model of NGC 1600. The right panels display, from top to bottom, images of the data, the best-fit model, and the residuals.  All images are shown using the same logarithmic stretch and are centered on the galaxy centroid, with north up and east to the left.  The left panels display the isophotal analysis of the 2D image fitting. From top to bottom, the panels show radial profiles of the the ellipticity ($e$), the position angle (PA), the $K_s$-band surface brightness ($\mu_{K_s}$), and the fitting residuals ($\bigtriangleup\mu_{K_s}$).  Profiles of the data, the model, and the individual components are encoded consistently with different symbols, line styles, and colors, which are explained in the legends.}
\figsetgrpend

\figsetgrpstart
\figsetgrpnum{37.9}
\figsetgrptitle{NGC 3091}
\figsetplot{NGC3091.pdf}
\figsetgrpnote{The best-fit three-component model of NGC 3091. The right panels display, from top to bottom, images of the data, the best-fit model, and the residuals.  All images are shown using the same logarithmic stretch and are centered on the galaxy centroid, with north up and east to the left.  The left panels display the isophotal analysis of the 2D image fitting. From top to bottom, the panels show radial profiles of the the ellipticity ($e$), the position angle (PA), the $K_s$-band surface brightness ($\mu_{K_s}$), and the fitting residuals ($\bigtriangleup\mu_{K_s}$).  Profiles of the data, the model, and the individual components are encoded consistently with different symbols, line styles, and colors, which are explained in the legends.}
\figsetgrpend

\figsetgrpstart
\figsetgrpnum{37.10}
\figsetgrptitle{NGC 3377}
\figsetplot{NGC3377.pdf}
\figsetgrpnote{The best-fit three-component model of NGC 3377. The right panels display, from top to bottom, images of the data, the best-fit model, and the residuals.  All images are shown using the same logarithmic stretch and are centered on the galaxy centroid, with north up and east to the left.  The left panels display the isophotal analysis of the 2D image fitting. From top to bottom, the panels show radial profiles of the the ellipticity ($e$), the position angle (PA), the $K_s$-band surface brightness ($\mu_{K_s}$), and the fitting residuals ($\bigtriangleup\mu_{K_s}$).  Profiles of the data, the model, and the individual components are encoded consistently with different symbols, line styles, and colors, which are explained in the legends.}
\figsetgrpend

\figsetgrpstart
\figsetgrpnum{37.11}
\figsetgrptitle{NGC 3379}
\figsetplot{NGC3379.pdf}
\figsetgrpnote{The best-fit three-component model of NGC 3379. The right panels display, from top to bottom, images of the data, the best-fit model, and the residuals.  All images are shown using the same logarithmic stretch and are centered on the galaxy centroid, with north up and east to the left.  The left panels display the isophotal analysis of the 2D image fitting. From top to bottom, the panels show radial profiles of the the ellipticity ($e$), the position angle (PA), the $K_s$-band surface brightness ($\mu_{K_s}$), and the fitting residuals ($\bigtriangleup\mu_{K_s}$).  Profiles of the data, the model, and the individual components are encoded consistently with different symbols, line styles, and colors, which are explained in the legends.}
\figsetgrpend

\figsetgrpstart
\figsetgrpnum{37.12}
\figsetgrptitle{NGC 3608}
\figsetplot{NGC3608.pdf}
\figsetgrpnote{The best-fit three-component model of NGC 3608. The right panels display, from top to bottom, images of the data, the best-fit model, and the residuals.  All images are shown using the same logarithmic stretch and are centered on the galaxy centroid, with north up and east to the left.  The left panels display the isophotal analysis of the 2D image fitting. From top to bottom, the panels show radial profiles of the the ellipticity ($e$), the position angle (PA), the $K_s$-band surface brightness ($\mu_{K_s}$), and the fitting residuals ($\bigtriangleup\mu_{K_s}$).  Profiles of the data, the model, and the individual components are encoded consistently with different symbols, line styles, and colors, which are explained in the legends.}
\figsetgrpend

\figsetgrpstart
\figsetgrpnum{37.13}
\figsetgrptitle{NGC 4374}
\figsetplot{NGC4374.pdf}
\figsetgrpnote{The best-fit three-component model of NGC 4374. The right panels display, from top to bottom, images of the data, the best-fit model, and the residuals.  All images are shown using the same logarithmic stretch and are centered on the galaxy centroid, with north up and east to the left.  The left panels display the isophotal analysis of the 2D image fitting. From top to bottom, the panels show radial profiles of the the ellipticity ($e$), the position angle (PA), the $K_s$-band surface brightness ($\mu_{K_s}$), and the fitting residuals ($\bigtriangleup\mu_{K_s}$).  Profiles of the data, the model, and the individual components are encoded consistently with different symbols, line styles, and colors, which are explained in the legends.}
\figsetgrpend

\figsetgrpstart
\figsetgrpnum{37.14}
\figsetgrptitle{NGC 4472}
\figsetplot{NGC4472.pdf}
\figsetgrpnote{The best-fit three-component model of NGC 4472. The right panels display, from top to bottom, images of the data, the best-fit model, and the residuals.  All images are shown using the same logarithmic stretch and are centered on the galaxy centroid, with north up and east to the left.  The left panels display the isophotal analysis of the 2D image fitting. From top to bottom, the panels show radial profiles of the the ellipticity ($e$), the position angle (PA), the $K_s$-band surface brightness ($\mu_{K_s}$), and the fitting residuals ($\bigtriangleup\mu_{K_s}$).  Profiles of the data, the model, and the individual components are encoded consistently with different symbols, line styles, and colors, which are explained in the legends.}
\figsetgrpend

\figsetgrpstart
\figsetgrpnum{37.15}
\figsetgrptitle{NGC 4473}
\figsetplot{NGC4473.pdf}
\figsetgrpnote{The best-fit three-component model of NGC 4473. The right panels display, from top to bottom, images of the data, the best-fit model, and the residuals.  All images are shown using the same logarithmic stretch and are centered on the galaxy centroid, with north up and east to the left.  The left panels display the isophotal analysis of the 2D image fitting. From top to bottom, the panels show radial profiles of the the ellipticity ($e$), the position angle (PA), the $K_s$-band surface brightness ($\mu_{K_s}$), and the fitting residuals ($\bigtriangleup\mu_{K_s}$).  Profiles of the data, the model, and the individual components are encoded consistently with different symbols, line styles, and colors, which are explained in the legends.}
\figsetgrpend

\figsetgrpstart
\figsetgrpnum{37.16}
\figsetgrptitle{M87}
\figsetplot{M87.pdf}
\figsetgrpnote{The best-fit three-component model of M87. The right panels display, from top to bottom, images of the data, the best-fit model, and the residuals.  All images are shown using the same logarithmic stretch and are centered on the galaxy centroid, with north up and east to the left.  The left panels display the isophotal analysis of the 2D image fitting. From top to bottom, the panels show radial profiles of the the ellipticity ($e$), the position angle (PA), the $K_s$-band surface brightness ($\mu_{K_s}$), and the fitting residuals ($\bigtriangleup\mu_{K_s}$).  Profiles of the data, the model, and the individual components are encoded consistently with different symbols, line styles, and colors, which are explained in the legends.}
\figsetgrpend

\figsetgrpstart
\figsetgrpnum{37.17}
\figsetgrptitle{NGC 4552}
\figsetplot{NGC4552.pdf}
\figsetgrpnote{The best-fit three-component model of NGC 4552. The right panels display, from top to bottom, images of the data, the best-fit model, and the residuals.  All images are shown using the same logarithmic stretch and are centered on the galaxy centroid, with north up and east to the left.  The left panels display the isophotal analysis of the 2D image fitting. From top to bottom, the panels show radial profiles of the the ellipticity ($e$), the position angle (PA), the $K_s$-band surface brightness ($\mu_{K_s}$), and the fitting residuals ($\bigtriangleup\mu_{K_s}$).  Profiles of the data, the model, and the individual components are encoded consistently with different symbols, line styles, and colors, which are explained in the legends.}
\figsetgrpend

\figsetgrpstart
\figsetgrpnum{37.18}
\figsetgrptitle{NGC 4621}
\figsetplot{NGC4621.pdf}
\figsetgrpnote{The best-fit three-component model of NGC 4621. The right panels display, from top to bottom, images of the data, the best-fit model, and the residuals.  All images are shown using the same logarithmic stretch and are centered on the galaxy centroid, with north up and east to the left.  The left panels display the isophotal analysis of the 2D image fitting. From top to bottom, the panels show radial profiles of the the ellipticity ($e$), the position angle (PA), the $K_s$-band surface brightness ($\mu_{K_s}$), and the fitting residuals ($\bigtriangleup\mu_{K_s}$).  Profiles of the data, the model, and the individual components are encoded consistently with different symbols, line styles, and colors, which are explained in the legends.}
\figsetgrpend

\figsetgrpstart
\figsetgrpnum{37.19}
\figsetgrptitle{NGC 4649}
\figsetplot{NGC4649.pdf}
\figsetgrpnote{The best-fit three-component model of NGC 4649. The right panels display, from top to bottom, images of the data, the best-fit model, and the residuals.  All images are shown using the same logarithmic stretch and are centered on the galaxy centroid, with north up and east to the left.  The left panels display the isophotal analysis of the 2D image fitting. From top to bottom, the panels show radial profiles of the the ellipticity ($e$), the position angle (PA), the $K_s$-band surface brightness ($\mu_{K_s}$), and the fitting residuals ($\bigtriangleup\mu_{K_s}$).  Profiles of the data, the model, and the individual components are encoded consistently with different symbols, line styles, and colors, which are explained in the legends.}
\figsetgrpend

\figsetgrpstart
\figsetgrpnum{37.20}
\figsetgrptitle{NGC 4889}
\figsetplot{NGC4889.pdf}
\figsetgrpnote{The best-fit three-component model of NGC 4889. The right panels display, from top to bottom, images of the data, the best-fit model, and the residuals.  All images are shown using the same logarithmic stretch and are centered on the galaxy centroid, with north up and east to the left.  The left panels display the isophotal analysis of the 2D image fitting. From top to bottom, the panels show radial profiles of the the ellipticity ($e$), the position angle (PA), the $K_s$-band surface brightness ($\mu_{K_s}$), and the fitting residuals ($\bigtriangleup\mu_{K_s}$).  Profiles of the data, the model, and the individual components are encoded consistently with different symbols, line styles, and colors, which are explained in the legends.}
\figsetgrpend

\figsetgrpstart
\figsetgrpnum{37.21}
\figsetgrptitle{NGC 5576}
\figsetplot{NGC5576.pdf}
\figsetgrpnote{The best-fit three-component model of NGC 5576. The right panels display, from top to bottom, images of the data, the best-fit model, and the residuals.  All images are shown using the same logarithmic stretch and are centered on the galaxy centroid, with north up and east to the left.  The left panels display the isophotal analysis of the 2D image fitting. From top to bottom, the panels show radial profiles of the the ellipticity ($e$), the position angle (PA), the $K_s$-band surface brightness ($\mu_{K_s}$), and the fitting residuals ($\bigtriangleup\mu_{K_s}$).  Profiles of the data, the model, and the individual components are encoded consistently with different symbols, line styles, and colors, which are explained in the legends.}
\figsetgrpend

\figsetgrpstart
\figsetgrpnum{37.22}
\figsetgrptitle{NGC 5813}
\figsetplot{NGC5813.pdf}
\figsetgrpnote{The best-fit three-component model of NGC 5813. The right panels display, from top to bottom, images of the data, the best-fit model, and the residuals.  All images are shown using the same logarithmic stretch and are centered on the galaxy centroid, with north up and east to the left.  The left panels display the isophotal analysis of the 2D image fitting. From top to bottom, the panels show radial profiles of the the ellipticity ($e$), the position angle (PA), the $K_s$-band surface brightness ($\mu_{K_s}$), and the fitting residuals ($\bigtriangleup\mu_{K_s}$).  Profiles of the data, the model, and the individual components are encoded consistently with different symbols, line styles, and colors, which are explained in the legends.}
\figsetgrpend

\figsetgrpstart
\figsetgrpnum{37.23}
\figsetgrptitle{NGC 5846}
\figsetplot{NGC5846.pdf}
\figsetgrpnote{The best-fit three-component model of NGC 5846. The right panels display, from top to bottom, images of the data, the best-fit model, and the residuals.  All images are shown using the same logarithmic stretch and are centered on the galaxy centroid, with north up and east to the left.  The left panels display the isophotal analysis of the 2D image fitting. From top to bottom, the panels show radial profiles of the the ellipticity ($e$), the position angle (PA), the $K_s$-band surface brightness ($\mu_{K_s}$), and the fitting residuals ($\bigtriangleup\mu_{K_s}$).  Profiles of the data, the model, and the individual components are encoded consistently with different symbols, line styles, and colors, which are explained in the legends.}
\figsetgrpend

\figsetgrpstart
\figsetgrpnum{37.24}
\figsetgrptitle{NGC 4697}
\figsetplot{NGC4697.pdf}
\figsetgrpnote{The best-fit four-component model of NGC 4697. The right panels display, from top to bottom, images of the data, the best-fit model, and the residuals.  All images are shown using the same logarithmic stretch and are centered on the galaxy centroid, with north up and east to the left.  The left panels display the isophotal analysis of the 2D image fitting. From top to bottom, the panels show radial profiles of the the ellipticity ($e$), the position angle (PA), the $K_s$-band surface brightness ($\mu_{K_s}$), and the fitting residuals ($\bigtriangleup\mu_{K_s}$).  Profiles of the data, the model, and the individual components are encoded consistently with different symbols, line styles, and colors, which are explained in the legends.}
\figsetgrpend

\figsetgrpstart
\figsetgrpnum{37.25}
\figsetgrptitle{NGC 3842}
\figsetplot{NGC3842.pdf}
\figsetgrpnote{The best-fit two-component model of NGC 3842. The right panels display, from top to bottom, images of the data, the best-fit model, and the residuals.  All images are shown using the same logarithmic stretch and are centered on the galaxy centroid, with north up and east to the left.  The left panels display the isophotal analysis of the 2D image fitting. From top to bottom, the panels show radial profiles of the the ellipticity ($e$), the position angle (PA), the $K_s$-band surface brightness ($\mu_{K_s}$), and the fitting residuals ($\bigtriangleup\mu_{K_s}$).  Profiles of the data, the model, and the individual components are encoded consistently with different symbols, line styles, and colors, which are explained in the legends.}
\figsetgrpend

\figsetgrpstart
\figsetgrpnum{37.26}
\figsetgrptitle{NGC 4291}
\figsetplot{NGC4291.pdf}
\figsetgrpnote{The best-fit two-component model of NGC 4291. The right panels display, from top to bottom, images of the data, the best-fit model, and the residuals.  All images are shown using the same logarithmic stretch and are centered on the galaxy centroid, with north up and east to the left.  The left panels display the isophotal analysis of the 2D image fitting. From top to bottom, the panels show radial profiles of the the ellipticity ($e$), the position angle (PA), the $K_s$-band surface brightness ($\mu_{K_s}$), and the fitting residuals ($\bigtriangleup\mu_{K_s}$).  Profiles of the data, the model, and the individual components are encoded consistently with different symbols, line styles, and colors, which are explained in the legends.}
\figsetgrpend

\figsetgrpstart
\figsetgrpnum{37.27}
\figsetgrptitle{NGC 4751}
\figsetplot{NGC4751.pdf}
\figsetgrpnote{The best-fit two-component model of NGC 4751. The right panels display, from top to bottom, images of the data, the best-fit model, and the residuals.  All images are shown using the same logarithmic stretch and are centered on the galaxy centroid, with north up and east to the left.  The left panels display the isophotal analysis of the 2D image fitting. From top to bottom, the panels show radial profiles of the the ellipticity ($e$), the position angle (PA), the $K_s$-band surface brightness ($\mu_{K_s}$), and the fitting residuals ($\bigtriangleup\mu_{K_s}$).  Profiles of the data, the model, and the individual components are encoded consistently with different symbols, line styles, and colors, which are explained in the legends.}
\figsetgrpend

\figsetgrpstart
\figsetgrpnum{37.28}
\figsetgrptitle{NGC 5077}
\figsetplot{NGC5077.pdf}
\figsetgrpnote{The best-fit two-component model of NGC 5077. The right panels display, from top to bottom, images of the data, the best-fit model, and the residuals.  All images are shown using the same logarithmic stretch and are centered on the galaxy centroid, with north up and east to the left.  The left panels display the isophotal analysis of the 2D image fitting. From top to bottom, the panels show radial profiles of the the ellipticity ($e$), the position angle (PA), the $K_s$-band surface brightness ($\mu_{K_s}$), and the fitting residuals ($\bigtriangleup\mu_{K_s}$).  Profiles of the data, the model, and the individual components are encoded consistently with different symbols, line styles, and colors, which are explained in the legends.}
\figsetgrpend

\figsetgrpstart
\figsetgrpnum{37.29}
\figsetgrptitle{NGC 5328}
\figsetplot{NGC5328.pdf}
\figsetgrpnote{The best-fit two-component model of NGC 5328. The right panels display, from top to bottom, images of the data, the best-fit model, and the residuals.  All images are shown using the same logarithmic stretch and are centered on the galaxy centroid, with north up and east to the left.  The left panels display the isophotal analysis of the 2D image fitting. From top to bottom, the panels show radial profiles of the the ellipticity ($e$), the position angle (PA), the $K_s$-band surface brightness ($\mu_{K_s}$), and the fitting residuals ($\bigtriangleup\mu_{K_s}$).  Profiles of the data, the model, and the individual components are encoded consistently with different symbols, line styles, and colors, which are explained in the legends.}
\figsetgrpend

\figsetgrpstart
\figsetgrpnum{37.30}
\figsetgrptitle{NGC 5516}
\figsetplot{NGC5516.pdf}
\figsetgrpnote{The best-fit two-component model of NGC 5516. The right panels display, from top to bottom, images of the data, the best-fit model, and the residuals.  All images are shown using the same logarithmic stretch and are centered on the galaxy centroid, with north up and east to the left.  The left panels display the isophotal analysis of the 2D image fitting. From top to bottom, the panels show radial profiles of the the ellipticity ($e$), the position angle (PA), the $K_s$-band surface brightness ($\mu_{K_s}$), and the fitting residuals ($\bigtriangleup\mu_{K_s}$).  Profiles of the data, the model, and the individual components are encoded consistently with different symbols, line styles, and colors, which are explained in the legends.}
\figsetgrpend

\figsetgrpstart
\figsetgrpnum{37.31}
\figsetgrptitle{NGC 5845}
\figsetplot{NGC5845.pdf}
\figsetgrpnote{The best-fit two-component model of NGC 5845. The right panels display, from top to bottom, images of the data, the best-fit model, and the residuals.  All images are shown using the same logarithmic stretch and are centered on the galaxy centroid, with north up and east to the left.  The left panels display the isophotal analysis of the 2D image fitting. From top to bottom, the panels show radial profiles of the the ellipticity ($e$), the position angle (PA), the $K_s$-band surface brightness ($\mu_{K_s}$), and the fitting residuals ($\bigtriangleup\mu_{K_s}$).  Profiles of the data, the model, and the individual components are encoded consistently with different symbols, line styles, and colors, which are explained in the legends.}
\figsetgrpend

\figsetgrpstart
\figsetgrpnum{37.32}
\figsetgrptitle{NGC 6086}
\figsetplot{NGC6086.pdf}
\figsetgrpnote{The best-fit two-component model of NGC 6086. The right panels display, from top to bottom, images of the data, the best-fit model, and the residuals.  All images are shown using the same logarithmic stretch and are centered on the galaxy centroid, with north up and east to the left.  The left panels display the isophotal analysis of the 2D image fitting. From top to bottom, the panels show radial profiles of the the ellipticity ($e$), the position angle (PA), the $K_s$-band surface brightness ($\mu_{K_s}$), and the fitting residuals ($\bigtriangleup\mu_{K_s}$).  Profiles of the data, the model, and the individual components are encoded consistently with different symbols, line styles, and colors, which are explained in the legends.}
\figsetgrpend

\figsetgrpstart
\figsetgrpnum{37.33}
\figsetgrptitle{NGC 6861}
\figsetplot{NGC6861.pdf}
\figsetgrpnote{The best-fit two-component model of NGC 6861. The right panels display, from top to bottom, images of the data, the best-fit model, and the residuals.  All images are shown using the same logarithmic stretch and are centered on the galaxy centroid, with north up and east to the left.  The left panels display the isophotal analysis of the 2D image fitting. From top to bottom, the panels show radial profiles of the the ellipticity ($e$), the position angle (PA), the $K_s$-band surface brightness ($\mu_{K_s}$), and the fitting residuals ($\bigtriangleup\mu_{K_s}$).  Profiles of the data, the model, and the individual components are encoded consistently with different symbols, line styles, and colors, which are explained in the legends.}
\figsetgrpend

\figsetgrpstart
\figsetgrpnum{37.34}
\figsetgrptitle{NGC 7619}
\figsetplot{NGC7619.pdf}
\figsetgrpnote{The best-fit two-component model of NGC 7619. The right panels display, from top to bottom, images of the data, the best-fit model, and the residuals.  All images are shown using the same logarithmic stretch and are centered on the galaxy centroid, with north up and east to the left.  The left panels display the isophotal analysis of the 2D image fitting. From top to bottom, the panels show radial profiles of the the ellipticity ($e$), the position angle (PA), the $K_s$-band surface brightness ($\mu_{K_s}$), and the fitting residuals ($\bigtriangleup\mu_{K_s}$).  Profiles of the data, the model, and the individual components are encoded consistently with different symbols, line styles, and colors, which are explained in the legends.}
\figsetgrpend

\figsetgrpstart
\figsetgrpnum{37.35}
\figsetgrptitle{NGC 7768}
\figsetplot{NGC7768.pdf}
\figsetgrpnote{The best-fit two-component model of NGC 7768. The right panels display, from top to bottom, images of the data, the best-fit model, and the residuals.  All images are shown using the same logarithmic stretch and are centered on the galaxy centroid, with north up and east to the left.  The left panels display the isophotal analysis of the 2D image fitting. From top to bottom, the panels show radial profiles of the the ellipticity ($e$), the position angle (PA), the $K_s$-band surface brightness ($\mu_{K_s}$), and the fitting residuals ($\bigtriangleup\mu_{K_s}$).  Profiles of the data, the model, and the individual components are encoded consistently with different symbols, line styles, and colors, which are explained in the legends.}
\figsetgrpend

\figsetgrpstart
\figsetgrpnum{37.36}
\figsetgrptitle{NGC 4486A}
\figsetplot{NGC4486A.pdf}
\figsetgrpnote{The best-fit single-component model of NGC 4486A. The right panels display, from top to bottom, images of the data, the best-fit model, and the residuals.  All images are shown using the same logarithmic stretch and are centered on the galaxy centroid, with north up and east to the left.  The left panels display the isophotal analysis of the 2D image fitting. From top to bottom, the panels show radial profiles of the the ellipticity ($e$), the position angle (PA), the $K_s$-band surface brightness ($\mu_{K_s}$), and the fitting residuals ($\bigtriangleup\mu_{K_s}$).  Profiles of the data, the model, and the individual components are encoded consistently with different symbols, line styles, and colors, which are explained in the legends.}
\figsetgrpend

\figsetgrpstart
\figsetgrpnum{37.37}
\figsetgrptitle{Holm 15A}
\figsetplot{Holm15A.pdf}
\figsetgrpnote{The best-fit single-component model of Holm 15A. The right panels display, from top to bottom, images of the data, the best-fit model, and the residuals.  All images are shown using the same logarithmic stretch and are centered on the galaxy centroid, with north up and east to the left.  The left panels display the isophotal analysis of the 2D image fitting. From top to bottom, the panels show radial profiles of the the ellipticity ($e$), the position angle (PA), the $K_s$-band surface brightness ($\mu_{K_s}$), and the fitting residuals ($\bigtriangleup\mu_{K_s}$).  Profiles of the data, the model, and the individual components are encoded consistently with different symbols, line styles, and colors, which are explained in the legends.}
\figsetgrpend

\figsetend

\clearpage

\startlongtable
\begin{deluxetable*}{cccrcrccc}
\tablecaption{Galaxy Sample and their Physical Properties\label{tab:1}}
\tablenum{1}
\tablehead{
\colhead{Galaxy} & \colhead{Hubble} & \colhead{$D_L$} & \colhead{$M_{\bullet}$(low $M_{\bullet}$--high $M_{\bullet}$)} & \colhead{$\sigma_\star$} & \colhead{log $M_{\text{bulge}}$} & \colhead{log $M_{\text{core}}$} & \colhead{Reference} & \colhead{{Class}} \\
\colhead{Name} & \colhead{Type} & \colhead{(Mpc)} & \colhead{($M_{\odot}$)} & \colhead{(km~s$^{-1}$)} & \colhead{($M_{\odot}$)} & \colhead{($M_{\odot}$)} & \colhead{} & \colhead{}}
\colnumbers
\startdata
Holm 15A & E & 244.0\tablenotemark{a} & $40(32\textendash48)\times10^{9}$ & 346$\pm$12 & 12.40$\pm$0.10 & \nodata\tablenotemark{b} & 4 & SR \\
IC 1459 & E4 & 28.92 & $2.48(2.29\textendash2.96)\times10^9$ & 331$\pm$5 & 11.60$\pm$0.09 & 11.26$\pm$0.09 & 1 & SR \\
M32 & E2 & 0.805 & $2.45(1.43\textendash3.46)\times10^6$ & 77$\pm$3 & 9.05$\pm$0.10 & 8.83$\pm$0.09 & 1 & FR \\
M87 & E1 & 16.68 & $6.15(5.78\textendash6.53)\times10^9$ & 324$\pm$28 & 11.72$\pm$0.09 & 11.16$\pm$ 0.15 & 1 & SR \\
NGC 1332 & E6 & 22.66 & $1.47(1.27\textendash1.68)\times10^9$ & 328$\pm$9 & 11.27$\pm$0.09 & 11.17$\pm$0.09 & 1 & FR \\
NGC 1374 & E0 & 19.57 & $5.90(5.39\textendash6.51)\times10^8$ & 167$\pm$3 & 10.65$\pm$0.09 & 10.50$\pm$0.10 & 1 & FR \\
NGC 1399 & E1 & 20.85 & $8.81(4.35\textendash17.81)\times10^8$ & 315$\pm$3 & 11.50$\pm$0.09 & 11.08$\pm$0.09 & 1 & SR \\
NGC 1407 & E0 & 29.00 & $4.65(4.24\textendash5.38)\times10^9$ & 276$\pm$2 & 11.74$\pm$0.09 & 11.39$\pm$0.09 & 1 & SR \\
NGC 1550 & E1 & 52.50 & $3.87(3.16\textendash4.48)\times10^9$ & 270$\pm$10 & 11.33$\pm$0.09 & 11.16$\pm$0.12 & 1 & SR \\
NGC 1600 & E3 & 64.0 & $1.7(1.55\textendash1.85)\times10^{10}$ & 293$\pm$15 & 11.92$\pm$0.09 & 11.41$\pm$0.09 & 3 & SR \\
NGC 2974 & E4 & 21.50 & $1.70(1.38\textendash2.10)\times10^8$ & 227$\pm$11 & 11.14$\pm$0.06 & \nodata\tablenotemark{c} & 2 & FR \\
NGC 3091 & E3 & 53.02 & $3.72(3.21\textendash3.83)\times10^9$ & 297$\pm$12 & 11.61$\pm$0.09 & 11.36$\pm$0.12 & 1 & SR \\
NGC 3377 & E5 & 10.99 & $1.78(0.85\textendash2.72)\times10^8$ & 145$\pm$7 & 10.50$\pm$0.09 & 10.28$\pm$0.09 & 1 & FR \\
NGC 3379 & E1 & 10.70 & $4.16(3.12\textendash5.20)\times10^8$ & 206$\pm$10 & 10.91$\pm$0.09 & 10.41$\pm$0.09 & 1 & SR \\
NGC 3608 & E1 & 22.75 & $4.65(3.66\textendash5.64)\times10^8$ & 182$\pm$9 & 11.01$\pm$0.09 & 10.63$\pm$0.11 & 1 & SR \\
NGC 3842 & E1 & 92.2 & $9.09(6.28\textendash11.43)\times10^9$ & 270$\pm$27 & 11.77$\pm$0.09 & 11.77$\pm$0.09 & 1 & SR \\
NGC 4291 & E2 & 26.58 & $9.78(6.70\textendash12.86)\times10^8$ & 242$\pm$12 & 10.85$\pm$0.09 & 10.40$\pm$0.09 & 1 & SR \\
NGC 4374 & E1 & 18.51 & $9.25(8.38\textendash10.23)\times10^8$ & 296$\pm$14 & 11.62$\pm$0.09 & 11.30$\pm$0.09 & 1 & SR \\
NGC 4472 & E2 & 16.72 & $2.54(2.44\textendash3.12)\times10^9$ & 300$\pm$7 & 11.84$\pm$0.09 & 11.26$\pm$0.09 & 1 & SR \\
NGC 4473 & E5 & 15.25 & $0.90(0.45\textendash1.35)\times10^8$ & 190$\pm$9 & 10.85$\pm$0.09 & 10.72$\pm$0.09 & 1 & FR \\
NGC 4486A & E2 & 18.36 & $1.44(0.92\textendash1.97)\times10^7$ & 111$\pm$5 & 10.04$\pm$0.09 & \nodata\tablenotemark{b} & 1 & FR \\
NGC 4552 & E & 15.30 & $5.01(4.47\textendash5.62)\times10^8$ & 252$\pm$12 & 11.42$\pm$0.11 & 10.73$\pm$0.09 & 2 & SR \\
NGC 4621 & E5 & 18.30 & $3.98(3.31\textendash4.79)\times10^8$ & 225$\pm$11 & 11.12$\pm$0.12 & 10.71$\pm$0.10 & 2 & FR \\
NGC 4649 & E2 & 16.46 & $4.72(3.67\textendash5.76)\times10^9$ & 380$\pm$19 & 11.64$\pm$0.09 & 11.22$\pm$0.09 & 1 & SR \\
NGC 4697 & E5 & 12.54 & $2.02(1.52\textendash2.53)\times10^8$ & 177$\pm$ 8 & 10.97$\pm$0.09 & 10.70$\pm$0.11 & 1 & N\\
NGC 4751 & E6 & 32.81 & $1.71(1.52\textendash1.81)\times10^9$ & 355$\pm$14 & 11.16$\pm$0.09 & 10.43$\pm$0.09 & 1 & FR \\
NGC 4889 & E4 & 102.0 & $20.8(4.9\textendash36.6)\times10^{9}$ & 347$\pm$5 & 12.09$\pm$0.09 & 11.89$\pm$0.09 & 1 & SR \\
NGC 5077 & E3 & 38.7 & $8.55(4.07\textendash12.93)\times10^8$ & 222$\pm$11 & 11.28$\pm$0.09 & 11.14$\pm$0.09 & 1 & SR \\
NGC 5328 & E2 & 64.4  & $4.75(2.81\textendash5.63)\times10^9$ & 333$\pm$2 & 11.65$\pm$0.09 & 11.10$\pm$0.09 & 1 & SR \\
%NGC 5419 & E & 56.20 & $7.24(5.25\textendash10.00)\times10^9$ & 367$\pm$10 & 12.01$\pm$0.15 & 11.73$\pm$0.09 & 2 & SR \\
NGC 5516 & E3 & 55.3 & $3.69(2.65\textendash3.79)\times10^9$ & 328$\pm$11 & 11.60$\pm$0.09 & 11.33$\pm$0.09 &1 & SR \\
NGC 5576 & E3 & 25.68 & $2.73(1.94\textendash3.41)\times10^8$ & 183$\pm$9 & 11.00$\pm$0.09 & 10.81$\pm$0.10 & 1 & FR \\
NGC 5813 & E1--2 & 32.20 & $7.08(6.17\textendash8.13)\times10^8$ & 230$\pm$12 & 11.77$\pm$0.09 & 11.02$\pm$0.09 & 2 & SR \\
NGC 5845 & E3 & 25.87 & $4.87(3.34\textendash6.40)\times10^8$ & 239$\pm$11 & 10.57$\pm$0.09 & 10.57$\pm$0.09 & 1 & FR \\
NGC 5846 & E0--1 & 24.90 & $1.10(0.95\textendash1.26)\times10^9$ & 237$\pm$12 & 11.62$\pm$0.16 & 11.17$\pm$0.09 & 2 & SR \\
NGC 6086 & E & 138.0 & $3.74(2.59\textendash5.50)\times10^9$ & 318$\pm$2 & 11.69$\pm$0.09 & 11.53$\pm$0.09 & 1 & SR \\
NGC 6861 & E4 & 28.71 & $2.10(2.00\textendash2.73)\times10^9$ & 389$\pm$3 & 11.25$\pm$0.09 & 11.15$\pm$0.09 & 1 & FR \\
NGC 7619 & E3 & 53.85 & $2.30(2.19\textendash3.45)\times10^9$ & 292$\pm$5 & 11.65$\pm$0.09 & 11.30$\pm$0.09& 1 & SR \\
NGC 7768 & E4 & 116.0 & $1.34(0.93\textendash1.85)\times10^9$ & 257$\pm$26 & 11.75$\pm$0.09 & 11.67$\pm$0.09 &1 & SR \\
M31      & Sb  & 0.774 & $1.43(1.12\textendash2.34)\times10^8$ & 169$\pm$8  & 10.35$\pm$0.09 & \nodata & 1 & \nodata \\
M81      & Sb  & 3.604 & $6.5(5.\textendash9.)\times10^7$ & 143$\pm$7  & 10.42$\pm$0.09 & \nodata & 1 & \nodata \\
NGC  524 & S0  & 24.22  & $8.67(8.21\textendash9.61)\times10^8$ & 247$\pm$12 & 11.26$\pm$0.09 & \nodata & 1 & \nodata \\
NGC  821 & S0  & 23.44 & $1.65(0.92\textendash2.39)\times10^8$ & 209$\pm$10 & 10.98$\pm$0.09 & \nodata & 1 & \nodata \\
NGC 1023 & SB0 & 10.81 & $4.13(3.71\textendash4.56)\times10^7$ & 205$\pm$10 & 10.53$\pm$0.09 & \nodata & 1 & \nodata \\
NGC 1194 & S0  & 57.98 & $7.08(6.76\textendash7.41)\times10^7$ & 148$\pm$24 & 10.64$\pm$0.09 & \nodata & 1 & \nodata \\
NGC 2549 & S0  & 12.70 & $1.45(0.31\textendash1.65)\times10^7$ & 145$\pm$7  &  9.71$\pm$0.09 & \nodata & 1 & \nodata \\
NGC 3115 & S0  &  9.54 & $8.97(6.20\textendash9.54)\times10^8$ & 230$\pm$11 & 10.92$\pm$0.09 & \nodata & 1 & \nodata \\
NGC 3245 & S0  & 21.38 & $2.39(1.63\textendash2.66)\times10^8$ & 205$\pm$10 & 10.69$\pm$0.09 & \nodata & 1 & \nodata \\
NGC 3585 & S0  & 20.51 & $3.29(2.71\textendash4.74)\times10^8$ & 213$\pm$11 & 11.26$\pm$0.09 & \nodata & 1 & \nodata \\
NGC 3998 & S0  & 14.30 & $8.45(7.79\textendash9.15)\times10^8$ & 275$\pm$7 & 10.67$\pm$0.09 & \nodata & 1  & \nodata \\
NGC 4026 & S0  & 13.35 & $1.80(1.45\textendash2.40)\times10^8$ & 180$\pm$9  & 10.33$\pm$0.09 & \nodata & 1 & \nodata \\
NGC 4258 & SABbc &  7.27 & $3.78(3.74\textendash3.82)\times10^7$ & 115$\pm$10 &9.86$\pm$0.09 & \nodata & 1 & \nodata \\
NGC 4526 & S0  & 16.44 & $4.51(3.48\textendash5.91)\times10^8$ & 222$\pm$11 & 11.02$\pm$0.09 & \nodata & 1 & \nodata \\
NGC 4564 & S0  & 15.94 & $8.81(6.38\textendash11.26)\times10^7$ & 162$\pm$8  & 10.38$\pm$0.09 & \nodata &1 & \nodata \\
NGC 4594 & Sa  &  9.87 & $6.65(6.24\textendash7.05)\times10^8$ & 240$\pm$12 & 11.47$\pm$0.09 & \nodata & 1 & \nodata \\
NGC 7457 & S0  & 12.53 & $0.90(0.36\textendash1.43)\times10^7$ &  67$\pm$3  &  9.56$\pm$0.09 & \nodata & 1 & \nodata \\
\enddata
\tablecomments{Column 1: galaxy name. Column 2: Hubble type. Column 3: luminosity distance. Column 4: BH mass. Column 5: stellar velocity dispersion of measured at $R_e$; in the case of disk galaxies, this pertains to the bulge alone. Column 6: total stellar mass, which for disk galaxies pertains to the bulge component alone. Column 7: core stellar mass, which, in the context of this study, applies only to the ellipticals; derived from the best-fit models in Table~\ref{tab:2}.  Column 8: literature reference: (1) \citetalias{KH13}; (2) \citealt{sa16}; (3) \citealt{tho16}; (4) \citealt{meh19}. { Column 9: class of elliptical galaxy: SR = slow rotator; FR = fast rotatior \citep{KH13,sa16,tho16,meh19}}.}
\tablenotetext{a}{Distance adjusted to our cosmology.}
\tablenotetext{b}{ Unreliable model because the compact angular size of the galaxy is poorly resolved (Appendix~\ref{sec:Appb}).}
\tablenotetext{c}{Unreliable model because of contamination by a nearby bright star (Section~\ref{sec:3.3}). }
\end{deluxetable*}

\startlongtable
\begin{deluxetable*}{ccccccccc}
\tablecaption{Parameters of Best-fit Models for Elliptical Galaxies \label{tab:2}}
\tablenum{2}
\tablehead{
\colhead{Galaxy} & \colhead{Model} & \colhead{$R_e$} & \colhead{$K_s$} & \colhead{$n$} & \colhead{$\epsilon$} & \colhead{PA} & \colhead{$f$} & \colhead{$\sigma_{\rm sky}$} \\
\colhead{} & \colhead{} & \colhead{(kpc)} & \colhead{(mag)} & \colhead{} & \colhead{} & \colhead{(deg)} & \colhead{} & \colhead{(counts)}}
\colnumbers
\startdata
Holm 15A & 1 & 20.4$\pm$0.3 & 9.94$\pm$0.01 & 1.34$\pm$0.01 & 0.27$\pm$0.01 & 155.8$\pm$0.9 & \nodata & 0.14\\
IC 1459 & 3 & 0.55$\pm$0.01 & 8.11$\pm$0.02 & 2.04$\pm$0.02 & 0.29$\pm$0.01 & 31.8$\pm$0.2 & 0.34$\pm$0.01 & 0.19 \\
  & & 2.01$\pm$0.05 & 8.8$\pm$0.2 & 0.90$\pm$0.05 & 0.32$\pm$0.02 & 49$\pm$1 & 0.17$\pm$0.02 & \\
  & & 5.12$\pm$0.02 & 7.69$\pm$0.04 & 0.71$\pm$0.02 & 0.25$\pm$0.01 & 38.2$\pm$0.4 & 0.49$\pm$0.01 & \\
M32 & 3 & 0.026$\pm$0.002 & 6.58$\pm$0.04 & 4.6$\pm$0.2 & 0.31$\pm$0.01 & 163.9$\pm$0.2 & 0.268$\pm$0.008 & 0.35 \\
  & & 0.093$\pm$0.002 & 6.00$\pm$0.03 & 1.26$\pm$0.05 & 0.22$\pm$0.02 & 155.8$\pm$0.3 & 0.457$\pm$0.007 & \\
  & & 0.205$\pm$0.004 & 6.55$\pm$0.04 & 0.69$\pm$0.08 & 0.11$\pm$0.01 & 178.7$\pm$0.5 & 0.275$\pm$0.007 & \\
M87 & 3 & 1.69$\pm$0.03 & 7.0$\pm$0.1 & 1.25$\pm$0.04 & 0.03$\pm$0.01 & 169$\pm$1 & 0.35$\pm$0.02 & 0.20\\
  & & 4.1$\pm$0.3 & 8.5$\pm$0.8 & 0.5$\pm$0.2 & 0.11$\pm$0.03 & 171$\pm$3 & 0.09$\pm$0.05 & \\
  & & 9$\pm$1 & 6.5$\pm$0.2 & 0.7$\pm$0.1 & 0.13$\pm$0.01 & 149$\pm$3 & 0.56$\pm$0.05 & \\
NGC 1332 & 3 & 0.553$\pm$0.004 & 8.16$\pm$0.01 & 2.04$\pm$0.02 & 0.26$\pm$0.01 & 119.1$\pm$0.2 & 0.384$\pm$0.003 & 0.19 \\
  & & 3.75$\pm$0.02 & 8.05$\pm$0.01 & 0.85$\pm$0.02 & 0.74$\pm$0.02 & 114.78$\pm$0.05 & 0.424$\pm$0.002 & \\
  & & 7.4$\pm$0.2 & 8.91$\pm$0.02 & 0.36$\pm$0.02 & 0.56$\pm$0.01 & 110.8$\pm$0.4 & 0.192$\pm$0.003 & \\
NGC 1374 & 3 & 0.35$\pm$0.07 & 9.4$\pm$0.2 & 3.8$\pm$0.4 & 0.10$\pm$0.01 & 114$\pm$2 & 0.35$\pm$0.04 & 0.14 \\ 
  & & 1.48$\pm$0.03 & 9.1$\pm$0.1 & 1.04$\pm$0.06 & 0.15$\pm$0.01 & 115$\pm$1 & 0.46$\pm$0.02 & \\
  & & 3.8$\pm$0.2 & 10.1$\pm$0.1 & 0.30$\pm$0.05 & 0.10$\pm$0.02 & 143$\pm$2 & 0.19$\pm$0.02 & \\
NGC 1399 & 3 & 0.344$\pm$0.002 & 9.20$\pm$0.02 & 0.98$\pm$0.01 & 0.18$\pm$0.01 & 99.1$\pm$0.8 & 0.102$\pm$0.005 & 0.29 \\
  & & 1.247$\pm$0.009 & 7.60$\pm$0.02 & 1.15$\pm$0.02 & 0.17$\pm$0.01 & 118.2$\pm$0.3 & 0.408$\pm$0.008 & \\
  & & 4.63$\pm$0.06 & 7.46$\pm$0.01 & 0.56$\pm$0.02 & 0.05$\pm$0.02 & 92$\pm$2 & 0.49$\pm$0.01 & \\
NGC 1407 & 3 & 0.653$\pm$0.009 & 8.69$\pm$0.02 & 1.18$\pm$0.01 & 0.04$\pm$0.01 & 49$\pm$1 & 0.21$\pm$0.01 & 0.26 \\
  & & 3.14$\pm$0.03 & 7.62$\pm$0.03 & 0.86$\pm$0.03 & 0.04$\pm$0.01 & 59$\pm$2 & 0.55$\pm$0.01 & \\
  & & 8.5$\pm$0.2 & 8.51$\pm$0.05 & 0.20$\pm$0.01 & 0.07$\pm$0.01 & 51$\pm$3 & 0.24$\pm$0.01 & \\
NGC 1550 & 3 & 1.3$\pm$0.1 & 9.7$\pm$0.1 & 3.5$\pm$0.2 & 0.30$\pm$0.01 & 33.0$\pm$0.5 & 0.44$\pm$0.02 & 0.14 \\
  & & 3.3$\pm$0.4 & 10.2$\pm$0.3 & 1.4$\pm$0.2 & 0.08$\pm$0.05 & 113$\pm$1 & 0.28$\pm$0.01 & \\
  & & 11$\pm$1 & 10.2$\pm$0.2 & 0.47$\pm$0.05 & 0.25$\pm$0.04 & 70$\pm$3 & 0.28$\pm$0.04 & \\
NGC 1600 & 3 & 2.14$\pm$0.03 & 9.53$\pm$0.02 & 1.18$\pm$0.01 & 0.35$\pm$0.01 & 6.4$\pm$0.3 & 0.24$\pm$0.02 & 0.15 \\
  & & 4.08$\pm$0.04 & 10.99$\pm$0.07 & 0.43$\pm$0.02 & 0.41$\pm$0.01 & 160$\pm$2 & 0.07$\pm$0.01 & \\
  & & 12.13$\pm$0.09 & 8.40$\pm$0.01 & 0.92$\pm$0.02 & 0.32$\pm$0.01 & 9.6$\pm$0.4 & 0.69$\pm$0.01 & \\
NGC 3091 & 3 & 0.54$\pm$0.04 & 10.8$\pm$0.3 & 1.18$\pm$0.05 & 0.34$\pm$0.01 & 135.0$\pm$0.1 & 0.09$\pm$0.02 & 0.14 \\
  & & 3.9$\pm$0.3 & 8.9$\pm$0.3 & 1.7$\pm$0.1 & 0.25$\pm$0.01 & 148.2$\pm$0.5 & 0.55$\pm$0.07 & \\
  & & 17$\pm$2 & 9.35$\pm$0.08 & 0.53$\pm$0.05 & 0.45$\pm$0.01 & 136.3$\pm$0.6 & 0.36$\pm$0.02 &\\
NGC 3377 & 3 & 0.38$\pm$0.03 & 8.56$\pm$0.05 & 4.3$\pm$0.2 & 0.53$\pm$0.01 & 43.1$\pm$0.1 & 0.35$\pm$0.01 & 0.14\\
  & & 1.87$\pm$0.01 & 8.45$\pm$0.04 & 1.03$\pm$0.03 & 0.49$\pm$0.01 & 41.4$\pm$0.1 & 0.39$\pm$0.01 & \\
  & & 4.2$\pm$0.1 & 8.89$\pm$0.02 & 0.44$\pm$0.01 & 0.24$\pm$0.01 & 48$\pm$1 & 0.26$\pm$0.01 & \\
NGC 3379 & 3 & 0.202$\pm$0.003 & 9.06$\pm$0.03 & 0.94$\pm$0.03 & 0.12$\pm$0.01 & 64$\pm$1 & 0.08$\pm$0.01 & 0.23\\
  & & 0.748$\pm$0.004 & 7.76$\pm$0.01 & 0.84$\pm$0.03 & 0.13$\pm$0.01 & 76.4$\pm$0.2 & 0.26$\pm$0.01& \\
  & & 3.71$\pm$0.02 & 6.77$\pm$0.01 & 0.90$\pm$0.02 & 0.14$\pm$0.01 & 67.67$\pm$0.2 & 0.66$\pm$0.01 &\\
NGC 3608 & 3 & 0.5$\pm$0.1 & 9.2$\pm$0.2 & 3.3$\pm$0.4 & 0.15$\pm$0.01 & 79.8$\pm$0.8 & 0.34$\pm$0.04 & 0.30\\
  & & 1.41$\pm$0.04 & 9.4$\pm$0.2 & 0.89$\pm$0.07 & 0.24$\pm$0.01 & 81.0$\pm$0.7 & 0.29$\pm$0.04 & \\
  & & 3.9$\pm$0.1 & 9.12$\pm$0.06& 0.38$\pm$0.02 & 0.24$\pm$0.01 & 75.2$\pm$0.6 & 0.37$\pm$0.01 & \\
NGC 3842 & 2 & 1.14$\pm$0.03 & 10.88$\pm$0.05 & 1.35$\pm$0.04 & 0.07$\pm$0.01 & 11.6$\pm$0.2 & 0.146$\pm$0.006 & 0.14 \\
  & & 11.6$\pm$0.2 & 8.96$\pm$0.02 & 1.78$\pm$0.06 & 0.24$\pm$0.01 & 175.8$\pm$0.7 & 0.854$\pm$0.002 &\\
NGC 4291 & 2 & 0.32$\pm$0.01 & 9.52$\pm$0.05 & 2.46$\pm$0.07 & 0.33$\pm$0.01 & 102.7$\pm$0.4 & 0.32$\pm$0.01 & 0.14\\
  & & 2.86$\pm$0.06 & 8.68$\pm$0.02 & 2.03$\pm$0.08 & 0.25$\pm$0.01 & 107.6$\pm$0.4 & 0.684$\pm$0.004 &\\
NGC 4374 & 3 & 0.628$\pm$0.007 & 7.74$\pm$0.02 & 1.58$\pm$0.03 & 0.20$\pm$0.01 & 124.7$\pm$0.2 & 0.25$\pm$0.04 & 0.29\\
  & & 2.89$\pm$0.05 & 7.00$\pm$0.02 & 1.00$\pm$0.03 & 0.10$\pm$0.01 & 125.7$\pm$0.4 & 0.502$\pm$0.005 & \\
  & & 6.95$\pm$0.08 & 7.78$\pm$0.04 & 0.33$\pm$0.03 & 0.02$\pm$0.01 & 133$\pm$9 & 0.244$\pm$0.006 & \\
NGC 4472 & 3 & 0.434$\pm$0.006 & 8.73$\pm$0.04 & 1.07$\pm$0.02 & 0.06$\pm$0.01 & 179$\pm$2 & 0.075$\pm$0.003 & 0.44 \\
  & & 2.25$\pm$0.02 & 6.59$\pm$0.02 & 1.15$\pm$0.04 & 0.17$\pm$0.01 & 161.0$\pm$0.2 & 0.54$\pm$0.04 & \\
  & & 6.90$\pm$0.06 & 6.96$\pm$0.02 & 0.30$\pm$0.02 & 0.19$\pm$0.01 & 156.4$\pm$0.3 & 0.38$\pm$0.04&\\
NGC 4473 & 3 & 1.09$\pm$0.01 & 8.19$\pm$0.02 & 1.94$\pm$0.02 & 0.42$\pm$0.01 & 93.81$\pm$0.07 & 0.385$\pm$0.005 & 0.16\\
  & & 5.2$\pm$0.1 & 8.23$\pm$0.05 & 1.04$\pm$0.04 & 0.47$\pm$0.01 & 93.0$\pm$0.1 & 0.37$\pm$0.01 & \\
  & & 12.1$\pm$0.2 & 8.68$\pm$0.07 & 0.42$\pm$0.03 & 0.39$\pm$0.01 & 91.8$\pm$0.4 & 0.25$\pm$0.01 & \\
NGC 4486A & 1 & 0.56$\pm$0.04 & 8.80$\pm$0.03 & 4.19$\pm$0.03 & 0.50$\pm$0.01 & 9.1$\pm$0.1 & \nodata & 0.20\\
NGC 4552 & 3 & 0.33$\pm$0.02 & 7.78$\pm$0.05 & 3.03$\pm$0.06 & 0.11$\pm$0.01 & 114.5$\pm$0.6 & 0.39$\pm$0.01& 0.18\\
  & & 0.348$\pm$0.003 & 9.73$\pm$0.04 & 0.51$\pm$0.02 & 0.17$\pm$0.01 & 34$\pm$1 & 0.065$\pm$0.003 &\\
  & & 1.63$\pm$0.04 & 7.41$\pm$0.03 & 1.07$\pm$0.02 & 0.09$\pm$0.01 & 129.8$\pm$0.9 & 0.546$\pm$0.007& \\
NGC 4621 & 3 & 0.42$\pm$0.08 & 9.3$\pm$0.2 & 6.9$\pm$0.4 & 0.65$\pm$0.03 & 164.4$\pm$0.3 & 0.12$\pm$0.02 & 0.24\\
  & & 0.56$\pm$0.05 & 8.3$\pm$0.1 & 2.5$\pm$0.2 & 0.22$\pm$0.02 & 162.8$\pm$0.6 & 0.29$\pm$0.02 &\\
  & & 2.61$\pm$0.08 & 7.53$\pm$0.05 & 1.17$\pm$0.03 & 0.42$\pm$0.02 & 164.1$\pm$0.2 & 0.59$\pm$0.01&\\
NGC 4649 & 3 & 0.368$\pm$0.002 & 9.51$\pm$0.03 & 0.71$\pm$0.02 & 0.17$\pm$0.01 & 118$\pm$1 & 0.041$\pm$0.001 & 0.40 \\
  & & 1.433$\pm$0.008 & 7.06$\pm$0.01 & 1.16$\pm$0.02 & 0.21$\pm$0.01 & 106.0$\pm$0.2 & 0.388$\pm$0.002&\\
  & & 5.81$\pm$0.04 & 6.64$\pm$0.02 & 0.58$\pm$0.02 & 0.20$\pm$0.01 & 101.8$\pm$0.2 & 0.571$\pm$0.004& \\
NGC 4697 & 4 & 0.9$\pm$0.2 & 8.3$\pm$0.2 & 3.1$\pm$0.2 & 0.43$\pm$0.01 & 65.6$\pm$0.2 & 0.22$\pm$0.03 & 0.25 \\
  & & 2.9$\pm$0.1 & 7.9$\pm$0.2 & 1.1$\pm$0.1 & 0.49$\pm$0.02 & 66.70$\pm$0.08 & 0.30$\pm$0.03 & \\
  & & 4.53$\pm$0.03 & 8.7$\pm$0.1 & 0.33$\pm$0.03 & 0.36$\pm$0.01 & 66.8$\pm$0.3 & 0.15$\pm$0.01 & \\
  & & 8.48$\pm$0.07 & 7.8$\pm$0.2 & 0.35$\pm$0.03 & 0.29$\pm$0.01 & 63.8$\pm$0.2 & 0.33$\pm$0.04 & \\
NGC 4751 & 2 & 0.38$\pm$0.02 & 10.03$\pm$0.07 & 2.6$\pm$0.1 & 0.41$\pm$0.01 & 173.9$\pm$0.5 & 0.19$\pm$0.01 & 0.16\\
  & & 4.3$\pm$0.1 & 8.47$\pm$0.02 & 2.64$\pm$0.06 & 0.60$\pm$0.01 & 174.70$\pm$0.01 & 0.807$\pm$0.002 &\\
NGC 4889 & 3 & 2.78$\pm$0.03 & 9.53$\pm$0.01 & 1.50$\pm$0.01 & 0.32$\pm$0.01 & 78.4$\pm$0.2 & 0.308$\pm$0.002 & 0.14\\
  & & 11.6$\pm$0.1 & 9.65$\pm$0.07 & 0.48$\pm$0.01 & 0.44$\pm$0.01 & 87.7$\pm$0.4 & 0.28$\pm$0.01 &\\
  & & 28$\pm$1 & 9.20$\pm$0.04 & 0.48$\pm$0.03 & 0.37$\pm$0.01 & 80.68$\pm$0.07 & 0.42$\pm$0.01&\\
NGC 5077 & 2 & 2.56$\pm$0.04 & 8.47$\pm$0.01 & 3.29$\pm$0.03 & 0.27$\pm$0.01 & 7.2$\pm$0.2 & 0.809$\pm$0.002 & 0.13\\
  & & 9.2$\pm$0.2 & 10.04$\pm$0.05 & 0.46$\pm$0.03 & 0.19$\pm$0.01 & 173.8$\pm$0.2 & 0.191$\pm$0.008 &\\
NGC 5328 & 2 & 1.34$\pm$0.05 & 9.73$\pm$0.05 & 1.98$\pm$0.04 & 0.40$\pm$0.01 & 88.5$\pm$0.3 & 0.31$\pm$0.01 & 0.16\\
  & & 8.5$\pm$0.2 & 8.85$\pm$0.02 & 1.46$\pm$0.05 & 0.29$\pm$0.01 & 82.2$\pm$0.7 & 0.692$\pm$0.004 &\\  
NGC 5516 & 2 & 2.65$\pm$0.09 & 8.88$\pm$0.03 & 3.22$\pm$0.06 & 0.16$\pm$0.01 & 162.2$\pm$0.6 & 0.514$\pm$0.007 & 0.19\\
  & & 15.0$\pm$0.2 & 8.94$\pm$0.02 & 0.79$\pm$0.03 & 0.28$\pm$0.01 & 156$\pm$1 & 0.486$\pm$0.004 &\\
NGC 5576 & 3 & 0.75$\pm$0.05 & 8.73$\pm$0.06 & 3.0$\pm$0.1 & 0.29$\pm$0.01 & 89.1$\pm$0.2 & 0.48$\pm$0.01 & 0.16\\
  & & 2.22$\pm$0.04 & 9.7$\pm$0.2 & 0.86$\pm$0.08 & 0.26$\pm$0.01 & 93$\pm$1 & 0.20$\pm$0.03 &\\
  & & 7.3$\pm$0.4 & 9.17$\pm$0.02 & 0.49$\pm$0.02 & 0.39$\pm$0.01 & 78.5$\pm$0.7 & 0.321$\pm$0.004 &\\
NGC 5813 & 3 & 0.64$\pm$0.03 & 9.21$\pm$0.05& 2.25$\pm$0.07 & 0.09$\pm$0.01 & 143$\pm$1 & 0.228$\pm$0.008 & 0.20\\
  & & 3.06$\pm$0.09 & 9.06$\pm$0.07 & 0.76$\pm$0.05 & 0.20$\pm$0.01 & 134$\pm$2 & 0.26$\pm$0.01 &\\
  & & 8.7$\pm$0.4 & 8.34$\pm$0.04 & 0.5$\pm$0.1 & 0.33$\pm$0.01 & 130.1$\pm$0.7 & 0.51$\pm$0.01&\\
NGC 5845 & 2 & 0.317$\pm$0.003 & 9.38$\pm$0.02 & 2.32$\pm$0.03 & 0.33$\pm$0.01 & 142.0$\pm$0.4 & 0.730$\pm$0.004 & 0.14\\
  & & 2.6$\pm$0.3 & 10.46$\pm$0.05 & 3.2$\pm$0.2 & 0.29$\pm$0.02 & 144$\pm$3 & 0.270$\pm$0.009 &\\
NGC 5846 & 3 & 1.14$\pm$0.06 & 8.64$\pm$0.07 & 1.86$\pm$0.05 & 0.11$\pm$0.01 & 91.2$\pm$0.9 & 0.25$\pm$0.01 & 0.23\\
  & & 3.64$\pm$0.04 & 7.90$\pm$0.05 & 0.83$\pm$0.03 & 0.08$\pm$0.01 & 33$\pm$2 & 0.49$\pm$0.01 & \\
  & & 9.4$\pm$0.2 & 8.60$\pm$0.05 & 0.21$\pm$0.01 & 0.08$\pm$0.01 & 68$\pm$4 & 0.26$\pm$0.01 & \\
NGC 6086 & 2 & 5.2$\pm$0.2 & 9.92$\pm$0.03 & 3.5$\pm$0.1 & 0.32$\pm$0.01 & 6.81$\pm$0.07 & 0.68$\pm$0.01 & 0.12\\
  & & 29.3$\pm$0.9 & 10.75$\pm$0.07 & 0.52$\pm$0.05 & 0.47$\pm$0.01 & 160$\pm$2 & 0.32$\pm$0.01 &\\
NGC 6861 & 2 & 1.83$\pm$0.01 & 7.96$\pm$0.01 & 2.60$\pm$0.03 & 0.47$\pm$0.01 & 142.8$\pm$0.07 & 0.770$\pm$0.002 & 0.14\\
  & & 8.1$\pm$0.1 & 9.27$\pm$0.03 & 0.59$\pm$0.04 & 0.25$\pm$0.01 & 130$\pm$2 & 0.230$\pm$0.005 &\\
NGC 7619 & 2 & 1.32$\pm$0.04 & 8.86$\pm$0.03 & 2.69$\pm$0.05 & 0.30$\pm$0.01 & 40.9$\pm$0.3 & 0.429$\pm$0.007 & 0.18\\
  & & 9.2$\pm$0.1 & 8.55$\pm$0.02 & 1.10$\pm$0.04 & 0.23$\pm$0.01 & 40.2$\pm$0.8 & 0.571$\pm$0.005 &\\
NGC 7768 & 2 & 4.7$\pm$0.2 & 9.57$\pm$0.03 & 2.80$\pm$0.07 & 0.26$\pm$0.01 & 57.3$\pm$0.7 & 0.652$\pm$0.007 & 0.16\\
  & & 23.6$\pm$0.6 & 10.25$\pm$0.06 & 0.49$\pm$0.03 & 0.37$\pm$0.01 & 49$\pm$2 & 0.35$\pm$0.01 & \\
\enddata
\tablecomments{Column 1: galaxy name. Column 2: number of model \sersic{} components used to fit the image. Column 3: effective radius. Column 4: apparent magnitude. Column 5: \sersic{} index. Column 6: ellipticity. Column 7: position angle. Column 8: light fraction of the component. Column 9: 1~$\sigma$ uncertainty of the sky. }
\end{deluxetable*}


\begin{thebibliography}{}

%\bibitem[Beifiori et al.(2012)]{bei12} Beifiori, A., Courteau, S., Corsini, E.~M., et al.\ 2012, \mnras, 419, 2497

%\bibitem[Bennert et al.(2011)]{ben11} Bennert, V.~N., Auger, M.~W., Treu, T., et al.\ 2011, \apj, 742, 107

\bibitem[Bentz \& Manne-Nicholas(2018)]{ben18} Bentz, M.~C., \& Manne-Nicholas, E.\ 2018, \apj, 864, 146

\bibitem[Bertin \& Arnouts(1996)]{bertin96} Bertin, E., \& Arnouts, S.\ 1996, \aaps, 117, 393

\bibitem[Bessell(2005)]{bes05} Bessell, M.~S.\ 2005, \araa, 43, 293

\bibitem[Bezanson et al.(2009)]{bez09} Bezanson, R., van Dokkum, P.~G., Tal, T., et al.\ 2009, \apj, 697, 1290

%\bibitem[Block et al.(2001)]{blo01} Block, D.~L., Puerari, I., Takamiya, M., et al.\ 2001, \aap, 371, 393

\bibitem[Bonfini et al.(2015)]{bon15} Bonfini, P., Dullo, B.~T., \& Graham, A.~W.\ 2015, \apj, 807, 136

\bibitem[Buitrago et al.(2008)]{buitrago08} Buitrago, F., Trujillo, I., Conselice, C. J., et al. 2008, \apj, 687, L61

\bibitem[Butterworth \& Harris(1992)]{but92} Butterworth, S.~T., \& Harris, W.~E.\ 1992, \aj, 103, 1828

\bibitem[Cappellari(2016)]{cap16} Cappellari, M.\ 2016, \araa, 54, 597

%\bibitem[Capaccioli et al.(2015)]{cap15} Capaccioli, M., Spavone, M., Grado, A., et al.\ 2015, \aap, 581, A10

\bibitem[Carpenter(2001)]{car01} Carpenter, J.~M.\ 2001, \aj, 121, 2851

\bibitem[Casali et al.(2007)]{cas07} Casali, M., Adamson, A., Alves de Oliveira, C., et al.\ 2007, \aap, 467, 777

\bibitem[Cole et al.(2020)]{col20} Cole, J., Bezanson, R., van der Wel, A., et al.\ 2020, \apjl, 890, L25

\bibitem[Cooper et al.(2013)]{coo13} Cooper, A.~P., D'Souza, R., Kauffmann, G., et al.\ 2013, \mnras, 434, 3348

\bibitem[Cooper et al.(2015)]{coo15} Cooper, A.~P., Parry, O.~H., Lowing, B., et al.\ 2015, \mnras, 454, 3185

\bibitem[Daddi et al.(2005)]{daddi05} Daddi, E., Renzini, A., Pirzkal, N., et al. 2005, \apj, 626, 680

\bibitem[Damjanov et al.(2011)]{dam11} Damjanov, I., Abraham, R. G., Glazebrook, K., et al. 2011, \apj, 739, L44

%\bibitem[Damjanov et al.(2009)]{dam09} Damjanov, I., McCarthy, P.~J., Abraham, R.~G., et al.\ 2009, \apj, 695, 101

\bibitem[Davari et al.(2017)]{davari17} Davari, R., Ho, L. C., Mobasher, B., \& Canalizo, G. 2017, \apj, 836, 75

\bibitem[Decarli et al.(2018)]{decarli18} Decarli, R., Walter, F., Venemans, B. P., et al. 2018, \apj, 855, 97

\bibitem[de la Rosa et al.(2016)]{del16} de la Rosa, I.~G., La Barbera, F., Ferreras, I., et al.\ 2016, \mnras, 457, 1916

\bibitem[Dom\'\i nguez S\'anchez et al.(2016)]{ds16} Dom\'\i nguez S\'anchez, H., P\'erez Gonz\'alez, P. G., Esquej, P., et al. 2016, \mnras, 457, 3743

\bibitem[D'Souza et al.(2014)]{dso14} D'Souza, R., Kauffman, G., Wang, J., et al.\ 2014, \mnras, 443, 1433

%\bibitem[Emsellem et al.(2007)]{emsellem07} Emsellem, E., Cappellari, M., Krajnovi\'c, D., et al. 2007, \mnras, 379, 401

\bibitem[Emsellem et al.(2011)]{emsellem11} Emsellem, E., Cappellari, M., Krajnovi\'c, D., et al. 2011, \mnras, 414, 888

\bibitem[Fan et al.(2008)]{fan08} Fan, L., Lapi, A., De Zotti, G., et al.\ 2008, \apjl, 689, L101

\bibitem[Ferrarese \& Merritt(2000)]{ferrarese00} Ferrarese, L., \& Merritt, D. 2000, \apj, 539, L9

\bibitem[Gabor \& Dav{\'e}(2012)]{gab12} Gabor, J.~M., \& Dav{\'e}, R.\ 2012, \mnras, 427, 1816

\bibitem[Gao \& Ho(2017)]{gao17} Gao, H., \& Ho, L.~C.\ 2017, \apj, 845, 114

\bibitem[Gebhardt et al.(2000)]{gebhardt00} Gebhardt, K., Bender, R., Bower, G., et al.  2000, \apj, 539, L13

%\bibitem[Genel et al.(2014)]{gen14} Genel, S., Vogelsberger, M., Springel, V., et al.\ 2014, \mnras, 445, 175

%%\bibitem[Graham \& Scott(2013)]{gra13} Graham, A.~W., \& Scott, N.\ 2013, \apj, 764, 151

%\bibitem[Graham \& Cappellari(2020a)]{gra19a} Graham, M.~T., \& Cappellari, M.\ 2020a, A\&A, submitted (arXiv:1910.05135)

%\bibitem[Graham et al.(2020b)]{gra19b} Graham, M.~T., Cappellari, M., Bershady, M.~A., et al.\ 2020b, A\&A, submitted (arXiv:1910.05136)

%\bibitem[Graham et al.(2020c)]{gra19c} Graham, M.~T., Cappellari, M., Bershady, M.~A., et al.\ 2020c, A\&A, submitted (arXiv:1910.05139)

\bibitem[Greene et al.(2010)]{gre10} Greene, J.~E., Peng, C.~Y., Kim, M., et al.\ 2010, \apj, 721, 26

\bibitem[Greene et al.(2020)]{gre20} Greene, J. E., Strader, J., \& Ho, L. C. 2020, ARA\&A, in press (arXiv:1911.09678)

\bibitem[Greene et al.(2019)]{gre19} Greene, J.~E., Veale, M., Ma, C.-P., et al.\ 2019, \apj, 874, 66

\bibitem[G{\"u}ltekin et al.(2009)]{gul09} G{\"u}ltekin, K., Richstone, D.~O., Gebhardt, K., et al.\ 2009, \apj, 698, 198

%\bibitem[Guo et al.(2011)]{guo11} Guo, Y., Giavalisco, M., Cassata, P., et al.\ 2011, \apj, 735, 18

\bibitem[Guo et al.(2009)]{guo09} Guo, Y., McIntosh, D.~H., Mo, H.~J., et al.\ 2009, \mnras, 398, 1129

\bibitem[H{\"a}ring \& Rix(2004)]{har04} H{\"a}ring, N., \& Rix, H.-W.\ 2004, \apjl, 604, L89

%\bibitem[Hirschmann et al.(2010)]{hir10} Hirschmann, M., Khochfar, S., Burkert, A., et al.\ 2010, \mnras, 407, 1016

\bibitem[Ho (2007)]{ho07} Ho, L. C. 2007, \apj, 669, 821

%\bibitem[Ho \& Kim(2014)]{ho14} Ho, L.~C., \& Kim, M.\ 2014, \apj, 789, 17

\bibitem[Ho et al.(2011)]{ho11} Ho, L.~C., Li, Z.-Y., Barth, A.~J., et al.\ 2011, \apjs, 197, 21

\bibitem[Hopkins et al.(2009)]{hopkins09} Hopkins, P. F., Bundy, K., Murray, N., et al. 2009, \mnras, 398, 898

%\bibitem[Hopkins et al.(2009)]{hop09} Hopkins, P.~F., Hernquist, L., Cox, T.~J., et al.\ 2009, \apj, 691, 1424

\bibitem[Hu (2008)]{hu08} Hu, J. 2008, \mnras, 386, 2242

\bibitem[Huang et al.(2013a)]{hu13} Huang, S., Ho, L.~C., Peng, C.~Y., et al.\ 2013a, \apj, 766, 47

\bibitem[Huang et al.(2013b)]{hu13b} Huang, S., Ho, L.~C., Peng, C.~Y., et al.\ 2013b, \apjl, 768, L28

\bibitem[Huang et al.(2016)]{hu16} Huang, S., Ho, L. C., Peng, C. Y., Li, Z.-Y., \& Barth, A. J. 2016, \apj, 821, 114

%\bibitem[Iodice et al.(2020)]{iod20} Iodice, E., Spavone, M., Cattapan, A., et al.\ 2020, \aap, 635, A3

\bibitem[Izumi et al.(2019)]{izumi19} Izumi, T., Onoue, M., Matsuoka, Y., et al. 2019, \pasj, 71, 111

\bibitem[Jahnke et al.(2009)]{jah09} Jahnke, K., Bongiorno, A., Brusa, M., et al.\ 2009, \apjl, 706, L215

%\bibitem[Jahnke \& Macci{\`o}(2011)]{jah11} Jahnke, K., \& Macci{\`o}, A.~V.\ 2011, \apj, 734, 92

\bibitem[Jarrett et al.(2000)]{jar00} Jarrett, T.~H., Chester, T., Cutri, R., et al.\ 2000, \aj, 119, 2498

\bibitem[Jarrett et al.(2003)]{jar03} Jarrett, T.~H., Chester, T., Cutri, R., et al.\ 2003, \aj, 125, 525

\bibitem[Jiang et al.(2012)]{jia12} Jiang, F., van Dokkum, P., Bezanson, R., et al.\ 2012, \apjl, 749, L10

\bibitem[Johansson et al.(2012)]{johansson12} Johansson, P. H., Naab, T., \& Ostriker, J. P. 2012, \apj, 754, 115

\bibitem[Kelly(2007)]{kel07} Kelly, B.~C.\ 2007, \apj, 665, 1489

%\bibitem[Kelvin et al.(2012)]{kel12} Kelvin, L.~S., Driver, S.~P., Robotham, A.~S.~G., et al.\ 2012, \mnras, 421, 1007

\bibitem[Khochfar et al.(2011)]{kho11} Khochfar, S., Emsellem, E., Serra, P., et al.\ 2011, \mnras, 417, 845

\bibitem[Komatsu et al.(2009)]{komatsu09} Komatsu, E., Dunkley, J., Nolta, M. R., et al. 2009, \apjs, 180, 330

\bibitem[Kormendy et al.(2011)]{kormendy11} Kormendy, J., Bender, R., \& Cornell, M. E. 2011, Nature, 469, 374

\bibitem[Kormendy et al.(2009)]{kor09} Kormendy, J., Fisher, D.~B., Cornell, M.~E., et al.\ 2009, \apjs, 182, 216

%\bibitem[Kormendy \& Gebhardt(2001)]{kor01} Kormendy, J., \& Gebhardt, K.\ 2001, in 20th Texas Symposium on Relativistic Astrophysics, ed. H. Martel \& J. C. Wheeler (Melville: AIP), 363

\bibitem[Kormendy \& Ho(2013)]{KH13} Kormendy, J., \& Ho, L.~C.\ 2013, \araa, 51, 511

\bibitem[Kormendy \& Kennicutt(2004)]{KormendyKennicutt2004} Kormendy, J., \& Kennicutt, R. C.  2004, \araa, 42, 603

\bibitem[Kormendy \& Richstone(1995)]{KR95}  Kormendy, J., \& Richstone, D. O. 1995, \araa, 33, 581

\bibitem[Kriek et al.(2008)]{Kriek08} Kriek, M., van der Wel, A., van Dokkum, P. G., Franx, M., \& Illingworth, G. D. 2008, \apj, 682, 896

\bibitem[Kriek et al.(2007)]{Kriek07} Kriek, M., van Dokkum, P. G., Franx, M., et al. 2007, \apj, 669, 776

\bibitem[Kron(1980)]{kro80} Kron, R.~G.\ 1980, \apjs, 43, 305

\bibitem[Labita et al.(2009)]{labita09} Labita, M., Decarli, R., Treves, A., \& Falomo, R. 2009, \mnras, 396, 1537

\bibitem[Li et al.(2011)]{li11} Li, Z.-Y., Ho, L.~C., Barth, A.~J., et al.\ 2011, \apjs, 197, 22

\bibitem[Liepold et al.(2020)]{Liepold20} Liepold, C. M., Quenneville, M. E., Ma, C.-P., et al. 2020, \apj, 891, 4

\bibitem[Magorrian et al.(1998)]{mag98} Magorrian, J., Tremaine, S., Richstone, D., et al.\ 1998, \aj, 115, 2285

%\bibitem[Maiolino et al.(2007)]{mai07} Maiolino, R., Neri, R., Beelen, A., et al.\ 2007, \aap, 472, L33

\bibitem[Marian et al.(2019)]{marian19} Marian, V., Jahnke, K., Mechtley, M., et al. 2019, \apj, 882, 141

\bibitem[McConnell \& Ma(2013)]{mcc13} McConnell, N.~J., \& Ma, C.-P.\ 2013, \apj, 764, 184

\bibitem[Mechtley et al.(2012)]{mechtley12} Mechtley, M., Windhorst, R. A., Ryan, R. E., et al. 2012, \apj, 756, L38

\bibitem[Mehrgan et al.(2019)]{meh19} Mehrgan, K., Thomas, J., Saglia, R., et al.\ 2019, ApJ, 887, 195

%\bibitem[Merloni et al.(2010)]{mer10} Merloni, A., Bongiorno, A., Bolzonella, M., et al.\ 2010, \apj, 708, 137

%\bibitem[Merritt \& Ferrarese(2001)]{mer01} Merritt, D., \& Ferrarese, L.\ 2001, \apj, 547, 140

\bibitem[Meyers (2015)]{mey15} Meyers, J.\ 2015, {\tt LinMix}, \url{https://github.com/jmeyers314/linmix}, GitHub

\bibitem[Moretti et al.(2014)]{mor14} Moretti, A., Poggianti, B.~M., Fasano, G., et al.\ 2014, \aap, 564, A138

\bibitem[Naab et al.(2007)]{na07} Naab, T., Johansson, P.~H., Ostriker, J.~P., et al.\ 2007, \apj, 658, 710

\bibitem[Naab et al.(2014)]{na14} Naab, T., Oser, L., Emsellem, E., et al.\ 2014, \mnras, 444, 3357

%\bibitem[Naab et al.(2009)]{na09} Naab, T., Johansson, P.~H., \& Ostriker, J.~P.\ 2009, \apjl, 699, L178

\bibitem[Oh et al.(2017)]{oh17} Oh, S., Greene, J.~E., \& Lackner, C.~N.\ 2017, \apj, 836, 115

%\bibitem[Oser et al.(2012)]{os12} Oser, L., Naab, T., Ostriker, J.~P., et al.\ 2012, \apj, 744, 63

\bibitem[Oser et al.(2010)]{os10} Oser, L., Ostriker, J.~P., Naab, T., et al.\ 2010, \apj, 725, 2312

\bibitem[Oyarz{\'u}n et al.(2019)]{oya19} Oyarz{\'u}n, G.~A., Bundy, K., Westfall, K.~B., et al.\ 2019, \apj, 880, 111

%\bibitem[Peng(2007)]{pen07} Peng, C.~Y.\ 2007, \apj, 671, 1098

\bibitem[Peng et al.(2002)]{pe02} Peng, C.~Y., Ho, L.~C., Impey, C.~D., et al.\ 2002, \aj, 124, 266

\bibitem[Peng et al.(2010)]{pe10} Peng, C.~Y., Ho, L.~C., Impey, C.~D., et al.\ 2010, \aj, 139, 2097

\bibitem[Peng et al.(2006)]{pen06} Peng, C.~Y., Impey, C.~D., Rix, H.-W., et al.\ 2006, \apj, 649, 616

\bibitem[Penoyre et al.(2017)]{pen17} Penoyre, Z., Moster, B.~P., Sijacki, D., et al.\ 2017, \mnras, 468, 3883

\bibitem[Pensabene et al.(2020)]{pensabene20} Pensabene, A., Carniani, S., Perna, M., et al. 2020, \aap, 637, A84

\bibitem[Pillepich et al.(2018)]{pil18} Pillepich, A., Nelson, D., Hernquist, L., et al.\ 2018, \mnras, 475, 648

\bibitem[Pulsoni et al.(2018)]{pul18} Pulsoni, C., Gerhard, O., Arnaboldi, M., et al.\ 2018, \aap, 618, A94

\bibitem[Pulsoni et al.(2020)]{pul20} Pulsoni, C., Gerhard, O., Arnaboldi, M., et al.\ 2020, \aap, 641, A60

%\bibitem[Reines \& Volonteri(2015)]{rei15} Reines, A.~E., \& Volonteri, M.\ 2015, \apj, 813, 82

\bibitem[Rodriguez-Gomez et al.(2016)]{rodriguez16} Rodriguez-Gomez, V., Pillepich, A., Sales, L. V., et al. 2016, \mnras, 458, 2371

\bibitem[Saglia et al.(2016)]{sa16} Saglia, R.~P., Opitsch, M., Erwin, P., et al.\ 2016, \apj, 818, 47

%\bibitem[Sani et al.(2011)]{san11} Sani, E., Marconi, A., Hunt, L.~K., et al.\ 2011, \mnras, 413, 1479

%\bibitem[Sarria et al.(2010)]{sar10} Sarria, J.~E., Maiolino, R., La Franca, F., et al.\ 2010, \aap, 522, L3

%\bibitem[Schramm \& Silverman(2013)]{sch13} Schramm, M., \& Silverman, J.~D.\ 2013, \apj, 767, 13

%\bibitem[Schutte et al.(2019)]{sch19} Schutte, Z., Reines, A., \& Greene, J.\ 2019, \apj, 887, 245

\bibitem[S\'ersic (1968)]{sersic68} S\'ersic, J. L. 1968, Atlas de Galaxias Australes (C\'ordoba: Obs. Astron., Univ. Nac. C\'ordoba)

\bibitem[Skrutskie et al.(2006)]{skr06} Skrutskie, M.~F., Cutri, R.~M., Stiening, R., et al.\ 2006, \aj, 131, 1163

\bibitem[Spavone et al.(2017)]{spa17} Spavone, M., Capaccioli, M., Napolitano, N.~R., et al.\ 2017, \aap, 603, A38

\bibitem[Sun et al.(2005)]{sun05} Sun, M., Vikhlinin, A., Forman, W., et al.\ 2005, \apj, 619, 169

%\bibitem[Targett et al.(2012)]{tar12} Targett, T.~A., Dunlop, J.~S., \& McLure, R.~J.\ 2012, \mnras, 420, 3621

\bibitem[Thomas et al.(2016)]{tho16} Thomas, J., Ma, C.-P., McConnell, N.~J., et al.\ 2016, \nat, 532, 340

\bibitem[Tokunaga et al.(2002)]{tok02} Tokunaga, A.~T., Simons, D.~A., \& Vacca, W.~D.\ 2002, \pasp, 114, 180

\bibitem[Tremaine et al.(2002)]{tremaine02} Tremaine, S., Gebhardt, K., Bender, R., et al. 2002, \apj, 574, 740

\bibitem[Trujillo et al.(2006)]{trujillo06} Trujillo, I., Feulner, G., Goranova, Y., et al. 2006, \mnras, 373, L36

\bibitem[Valentinuzzi et al.(2009)]{val09} Valentinuzzi, T., Woods, D., Fasano, G., et al.\ 2009, \aap, 501, 851

\bibitem[van der Wel et al.(2009)]{van09} van der Wel, A., Bell, E.~F., van den Bosch, F.~C., et al.\ 2009, \apj, 698, 1232

\bibitem[van der Wel et al.(2014)]{van14} van der Wel, A., Franx, M., van Dokkum, P.~G., et al.\ 2014, \apj, 788, 28

%\bibitem[van Dokkum et al.(2008)]{van08} van Dokkum, P. G., Franx, M., Kriek, M., et al. 2008, \apj, 677, L5

\bibitem[van Dokkum et al.(2010)]{van10} van Dokkum, P. G., Whitaker, K. E., Brammer, G., et al. 2010, \apj, 709, 1018

%\bibitem[Vogelsberger et al.(2014)]{vog14} Vogelsberger, M., Genel, S., Springel, V., et al.\ 2014, \mnras, 444, 1518

%\bibitem[Walter et al.(2004)]{wal04} Walter, F., Carilli, C., Bertoldi, F., et al.\ 2004, \apjl, 615, L17

\bibitem[Wang et al.(2010)]{wan10} Wang, R., Carilli, C.~L., Neri, R., et al.\ 2010, \apj, 714, 699

\bibitem[Wellons et al.(2016)]{wel16} Wellons, S., Torrey, P., Ma, C.-P., et al.\ 2016, \mnras, 456, 1030

%\bibitem[Wyithe(2006a)]{wyi06a} Wyithe, J.~S.~B.\ 2006a, \mnras, 365, 1082

%\bibitem[Wyithe(2006b)]{wyi06b} Wyithe, J.~S.~B.\ 2006b, \mnras, 371, 1536

\bibitem[Yoon et al.(2011)]{yoo11} Yoon, I., Weinberg, M.~D., \& Katz, N.\ 2011, \mnras, 414, 1625

\bibitem[Yu et al.(2018)]{yu18} Yu, S.-Y., Ho, L.~C., Barth, A.~J., \& Li, Z.-Y. 2018, \apj, 862, 13

%\bibitem[Zibetti et al.(2009)]{zib09} Zibetti, S., Charlot, S., \& Rix, H.-W.\ 2009, \mnras, 400, 1181

\end{thebibliography}
\end{document}